\documentclass[11pt,a4paper]{article}                           

\usepackage{amsmath}
\usepackage{amssymb}
\usepackage[usenames]{color} 
\usepackage{multirow}
\usepackage{graphicx}
\usepackage{epstopdf}
\usepackage{array}
\usepackage{hhline}
\usepackage{cite}
\usepackage{microtype,colortbl}
\usepackage[bf]{caption}
\usepackage{color}
\usepackage[usenames,dvipsnames]{xcolor}
\usepackage{pstool}
\usepackage{tikz}
\usepackage{ytableau}
\usepackage{latexsym,stmaryrd}
\usepackage[bookmarks=false]{hyperref}
 \usepackage{lscape}
 \usepackage{subfig}
 \usepackage{tensor}
 \usepackage{dsfont}
 \usepackage{tcolorbox}
 \usepackage{tikz-cd}
\usepackage{ifpdf}
\def\hybrid{\topmargin -20pt    \oddsidemargin 0pt
        \headheight 0pt \headsep 0pt
        \textwidth 6.25in       % A4 paper
        \textheight 9 in       % A4 paper
        \marginparwidth .875in
        \parskip 5pt plus 1pt 
          \jot = 1.5ex
   }
\hybrid
\numberwithin{equation}{section}
\numberwithin{table}{section}

\newcommand{\beq}{\begin{equation}}  \newcommand{\eeq}{\end{equation}}
\newcommand{\bal}{\begin{aligned}}   \newcommand{\eal}{\end{aligned}}
\newcommand{\bea}{\begin{eqnarray}}  \newcommand{\eea}{\end{eqnarray}}

\newcommand{\bmat}{\left(\begin{array}}
\newcommand{\emat}{\end{array}\right)}

\usepackage{cancel}

\newcommand{\be}{\begin{equation}}
\newcommand{\ee}{\end{equation}}

\newcommand{\dd}{\mathrm{d}}

\definecolor{Gray}{gray}{0.95}

\begin{document}
\baselineskip=14pt
\parskip 5pt plus 1pt 

\vspace*{2cm}
\begin{center}
	{\LARGE\bfseries Deformed WZW Models and Hodge Theory}\\[.3cm]
	{\Large \bfseries -- Part I -- }
	
	\vspace{1cm}
	{\bf Thomas W.~Grimm}\footnote{t.w.grimm@uu.nl},
	{\bf Jeroen Monnee}\footnote{j.monnee@uu.nl},
	
	{\small
		\vspace*{.5cm}
		Institute for Theoretical Physics, Utrecht University\\ Princetonplein 5, 3584 CC Utrecht, The Netherlands\\[3mm]
	}
\end{center}
\vspace{1.5cm}
\begin{abstract}\noindent
We investigate a relationship between a particular class of two-dimensional integrable non-linear $\sigma$-models and variations of Hodge structures. Concretely, our aim is to study the classical dynamics of the $\lambda$-deformed $G/G$ model and show that a special class of solutions to its equations of motion precisely describes a one-parameter variation of Hodge structures. We find that this special class is obtained by identifying the group-valued field of the $\sigma$-model with the Weil operator of the Hodge structure. In this way, the study of strings on classifying spaces of Hodge structures suggests an interesting connection between the broad field of integrable models and the mathematical study of period mappings.
\end{abstract}

\newpage

\tableofcontents
\setcounter{footnote}{0}

\newpage
%%%%%%%%%%%%%%%%%%%%%%%%%%%%%%%%%%%%%%%%%%%%%%%%%%%%%%

\section{Introduction}

Across numerous branches of physics, ranging from string theory to condensed matter physics, integrable non-linear $\sigma$-models have played a major role. 
Such models possess a large amount of symmetries and are constrained to the extent that they are often solvable, for example, via the Bethe ansatz. Nevertheless, despite these constraints there has emerged a vast and intricate network of different integrable models, many of which are obtained by suitably deforming known models such as the Wess-Zumino-Witten (WZW) model and its gauged extension \cite{Wess:1971,Witten:1983_current_algebra,Witten:1983_bosonization,Gawedzki:1988_I,Gawedzki:1988_II,Witten:1991}. At the same time, it has recently been suggested to study non-linear $\sigma$-models in the context of string moduli spaces \cite{Grimm:2020cda,Grimm:2021ikg}, see also \cite{Cecotti:2020uek} for related ideas. These works propose to use an auxiliary field theory on the moduli space to provide a physical reformulation of the intricate mathematical structures that arise, as described by (asymptotic) Hodge theory. This has motivated us to further develop a possible relationship between integrable models and the physics of string compactifications via variations of Hodge structures. 

In the set-ups of \cite{Grimm:2020cda,Cecotti:2020uek,Grimm:2021ikg}, the spacetime of the model is identified with the field space of the effective theory obtained from string compactification. The goal, then, is to study the properties of the effective theory by analysing the solutions to said model, with an appropriate set of boundary conditions. Concrete examples of such properties are the form of the gauge couplings in four-dimensional $\mathcal{N}=2$ supergravity theories coming from type IIB compactifications, as well as the scalar potential in four-dimensional $\mathcal{N}=1$ supergravity theories coming from flux compactifications of F-theory. The aim of this approach is to provide a new perspective to study some aspects of effective field theories, which may be more familiar to physicists. However, there still remains much to understand about the precise formulation of the non-linear $\sigma$-model. For example, restricting to two-dimensional field spaces, one might expect the model to be among the plethora of two-dimensional integrable field theories that have been studied extensively over the last decennia. It is the main purpose of this work to identify the appropriate field theory as an integrable deformation of the gauged WZW model. This results in the exciting possibility to use techniques of integrability to study the field spaces of string compactifications and, conversely, to use existing methods of %e.g. asymptotic 
Hodge theory and the study of period mappings to obtain explicit solutions to integrable field theories.

Concretely, we aim to establish a relationship between a particular class of integrable deformations of the gauged WZW model and the mathematical concept of a variation of Hodge structures (VHS) along a complex one-dimensional moduli space. In the past years, the study of variations of Hodge structures and in particular their asymptotic behaviour near the boundaries of field space have been used intensely to study string compactifications and to provide evidence for swampland conjectures stating general constraints on effective theories to 
be compatible with quantum gravity \cite{GPV,Grimm:2018cpv,Palti:2019pca,Grimm:2019ixq,Grimm:2019wtx,Font:2019cxq,Gendler:2020dfp,Lanza:2020qmt,Calderon-Infante:2020dhm,Grimm:2020cda,Bastian:2020egp,Grimm:2020ouv,Bastian:2021hpc,Grimm:2021ckh,Palti:2021ubp}. Intuitively, a VHS is nothing but a decomposition of a vector space into various pieces, with the `angle' between different constituents being parametrized by a complex modulus $t$. The prime example, and main motivation to study such objects, is given by the primitive middle cohomology of a K\"ahler manifold $X$, for which the decomposition in terms of $(p,q)$-forms depends on the complex structure of $X$. Naturally, the dependence of this decomposition on the modulus $t$ is not arbitrary and must satisfy the so-called horizontality condition. This can be conveniently described via the use of a grading element $Q(t,\bar{t})$, which we refer to as the `charge operator', for which the horizontality condition reduces to a differential equation. The crucial insight, then, is to reinterpret the horizontality condition as arising from the equations of motion of an integrable deformation of a two-dimensional $\sigma$-model.

Historically, one of the motivations for studying integrable deformations of two-dimensional $\sigma$-models has been to find a worldsheet action principle that describes the $q$-deformation of the S-matrix of the $AdS_5\times S^5$ superstring \cite{Hoare:2011wr,Delduc:2013qra,Delduc:2014kha}. Such deformations are of interest, since they reduce the amount of supersymmetry while retaining integrability, allowing for a more general study of the AdS/CFT correspondence beyond $\mathcal{N}=4$ SYM \cite{Borsato:2013qpa,Hoare:2015kla,Pachol:2015mfa}, see also \cite{Beisert:2010jr} for an elaborate list of references. Nevertheless, even in the bosonic setting integrable deformations of two-dimensional $\sigma$-models have received much attention \cite{Klimcik:2002zj,Klimcik:2008eq,Delduc:2013fga,Kawaguchi:2013gma,Sfetsos:2014_integrability,Hollowood:2014,Hollowood:2014qma,Hoare:2014pna,Klimcik:2014bta,Delduc:2014uaa,Kawaguchi:2014qwa,vanTongeren:2015uha,Osten:2016dvf} and a vast web of connections and dualities between them has been uncovered over the years \cite{Klimcik_2015,Klimcik:2016rov,Hoare:2016wsk,Borsato:2016pas,Borsato:2017qsx}, see also \cite{Georgiou:2021pbd,Klimcik:2021bjy,Hoare:2021dix} for recent reviews.

The precise deformations we will consider in this work are the so-called $\lambda$-deformations, first discovered by Sfetsos in \cite{Sfetsos:2014_integrability} for the principal chiral model and WZW model, and later generalized to symmetric and semi-symmetric spaces in \cite{Hollowood:2014,Hollowood:2014qma}. In their simplest form, one can view these as a one-parameter family of integrable $\sigma$-models that interpolate between the WZW model and the non-Abelian T-dual of the principal chiral model. Also the $\lambda$-deformations have seen considerable development and generalizations to e.g.~multi-parameter/asymmetric deformations \cite{Sfetsos:2015nya,Georgiou:2016urf,Georgiou:2018gpe,Driezen:2019ykp}. Additionally, recently great progress has been made on formulating integrable $\lambda$-deformations on worldsheets with boundaries and studying D-brane configurations, see e.g.~\cite{Alekseev:1998mc,Felder:1999ka,Figueroa-OFarrill:1999cmq,Driezen:2018glg,Sfetsos:2021pcs}. However, for the purpose of this work we will restrict our attention to the so-called $\lambda$-deformed $G/G$ model, which corresponds to the $\lambda$-deformation of the fully gauged WZW model. Its field content consists of a group-valued field and a gauge field. By appropriately identifying these fields with the aforementioned charge operator $Q(t,\bar{t})$ we show that the resulting equations of motion precisely describe a VHS. Conversely, this implies that any one-parameter VHS yields a solution to the $\lambda$-deformed $G/G$ model. In this first part, we will not be concerned with the boundary conditions of the fields, but leave this and its connection with asymptotic Hodge theory to a second part.

The paper is organized as follows. In section \ref{sec:lambda_WZW} we present a basic review of the gauged WZW model and its $\lambda$-deformation and study its classical dynamics via the equations of motion. In section \ref{sec:Hodge_theory} the notion of a VHS is discussed via the period mapping and the charge operator. Finally, in section \ref{sec:strings_Hodge_theory} the connection between the $\lambda$-deformed $G/G$ model and the notion of a VHS is made precise, by showing that a VHS realizes a particular solution to the equations of motion of the $\lambda$-deformed $G/G$ model. We also make some additional comments regarding integrability and provide an in-depth analysis of more general solutions at the end of the section. Finally, there are three appendices which contain some computational details.

\section{$\lambda$-deformed WZW models}
\label{sec:lambda_WZW}

In this section we provide a classical analysis of the $\lambda$-deformed $G/G$ model, with emphasis on its equations of motion. This model can be viewed as a deformation of the well-known $G/G$ model by a term in the action that explicitly breaks the gauge symmetry. We begin in section \ref{subsec:gWZW} with a standard review on the WZW model and the $G/G$ model. The $\lambda$-deformation of the latter is then discussed in section \ref{subsec:lambda_G/G}, where a detailed analysis of the resulting equations of motion is given. The important equations that will be used in subsequent sections are \eqref{eq:A_ginvdg}, \eqref{eq:Abar_ginvdbarg} and \eqref{eq:eomg_full}, together with \eqref{eq:starA_closed}. 

\subsection{Gauged WZW models}
\label{subsec:gWZW}

We start by reviewing classical aspects of the WZW model and its gauged extension, mostly to set the notation and fix our conventions. For an in-depth discussion on WZW models, also beyond the classical description, we refer the reader to e.g.~\cite{Gawedzki_1999}.

A WZW model is a non-linear $\sigma$-model defined on a two-dimensional worldsheet $\Sigma$ with a Lie group $G$ as target space, which we assume to be semi-simple. It describes the dynamics of a group-valued field 
\begin{equation}
	g: \Sigma\rightarrow G\,,
\end{equation}
whose action is given by the sum of the principal chiral model and the celebrated Wess-Zumino action  \cite{Wess:1971,Witten:1983_current_algebra,Witten:1983_bosonization}
\begin{equation}
\label{eq:WZW_action}
	S_{\mathrm{WZW}}[g]= \frac{k}{8\pi} \int_\Sigma \mathrm{Tr}\left(g^{-1}\dd g\wedge\star\, g^{-1}\dd g \right) + \frac{k}{12\pi i}\int \mathrm{Tr}\left(g^{-1}\dd g\wedge g^{-1}\dd g\wedge g^{-1}\dd g  \right)\,.
\end{equation}
Here $\star$ denotes the Hodge star on $\Sigma$ and the second term features an integration of a suitable extension of $g$ over a three-manifold whose boundary is the worldsheet $\Sigma$. Furthermore, $\mathrm{Tr}$ denotes any non-degenerate $\mathrm{Ad}$-invariant bilinear form on the Lie algebra $\mathfrak{g}$ of $G$. Finally, the constant $k$ is referred to as the level of the model. It is restricted to take integer values when $G$ is compact so that the path integral is well-defined. In the current work however, $k$ will not play an important role as we are merely concerned with classical features. 

By virtue of the trace, the action \eqref{eq:WZW_action} enjoys a global $G\times G$ symmetry 
\begin{equation}
\label{eq:global_symmetry}
	g\mapsto g_L\cdot g\cdot g_R^{-1}\,,\qquad (g_L, g_R) \in G\times G\,.
\end{equation}
One may now proceed to define a gauged WZW model by gauging a particular subgroup of this global symmetry, as reviewed in e.g.~\cite{Chung:1993}. In the current work, we will consider the vector gauged WZW model in which the diagonal subgroup of $G\times G$ is gauged, corresponding to the transformation $g\mapsto hgh^{-1}$, for $h\in G$. The resulting action is most conveniently written down using local coordinates $x,y$ on $\Sigma$, which we take have Euclidean signature, and passing to complex coordinates $t,\bar{t}$ via $t=x+iy$. Furthermore, by a suitable conformal transformation the metric on $\Sigma$ can be taken to be the flat metric, i.e.
\begin{equation}
	ds^2 = \dd x^2+\dd y^2 = \dd t\,\dd \bar{t}\,.
\end{equation}
Finally, the gauge field will be denoted by
\begin{equation}
	\mathbf{A} = A\,\dd t+\bar{A}\,\dd\bar{t}\,,
\end{equation}
and its components $A,\bar{A}$ are fields on $\Sigma$ taking values in $\mathfrak{g}$. Then the action of the vector gauged WZW model reads\footnote{Here $d^2t = \frac{i}{2}\dd t\wedge \dd \bar{t}=\dd x\wedge \dd y$ and furthermore $\partial = \partial_t$, $\bar{\partial} = \partial_{\bar{t}}$.} \cite{Gawedzki:1988_I,Gawedzki:1988_II,Witten:1991}
\begin{equation}
\label{eq:gWZW_action}
S_{G/G}[g,\mathbf{A}] = S_{\mathrm{WZW}}[g]+\frac{k}{\pi}\int_\Sigma d^2t\;\mathrm{Tr}\left(A\Bar{\partial}g g^{-1}-\bar{A}g^{-1}\partial g -Ag\Bar{A}g^{-1}+A\Bar{A} \right)\,.
\end{equation}
Indeed, one may verify that \eqref{eq:gWZW_action} is invariant under the gauge transformation
\begin{equation}
	g\mapsto hgh^{-1}\,,\qquad \mathbf{A}\mapsto h\left(\dd+\mathbf{A}\right)h^{-1}\,,\qquad h\in G\,.
\end{equation}
Since the resulting action is gauge-invariant under conjugation by the full group $G$, this model is also referred to as the $G/G$ model. For completeness and later reference, we record the equations of motion 
\begin{align}
\label{eq:eom_gWZW_barA}	&\delta\bar{A}:\qquad g^{-1}Dg = 0\,, \\
	&\delta A:\qquad \bar{D}g g^{-1}=0\,, \\
\label{eq:eom_gWZW_g}	&\delta g:\qquad F-\bar{D}(g^{-1}Dg)=0\,, 
\end{align}
which are obtained from the variation of \eqref{eq:gWZW_action} w.r.t. the various fields. Here $D,\bar{D}$ denote the covariant derivatives, defined by
\begin{equation}
	D = \partial + [A,-]\,,\qquad \bar{D} = \bar{\partial} + [\bar{A},-]\,,
\end{equation}
and $F$ denotes the field-strength, which is given by 
\begin{equation}
	F = \partial \bar{A} - \bar{\partial}A + [A,\bar{A}]\,.
\end{equation}
An important property of the $G/G$ model is the on-shell vanishing of the field strength, which is easily seen by combining \eqref{eq:eom_gWZW_barA} and \eqref{eq:eom_gWZW_g}. As a result, the gauge field is (locally) pure-gauge and hence it does not contain any physical degrees of freedom. As will become apparent, this is no longer true for the $\lambda$-deformed $G/G$ model.

\subsection{$\lambda$-deformations}
\label{subsec:lambda_G/G}

Let us now introduce the $\lambda$-deformed $G/G$ model, following the discussion in \cite{Hollowood:2014}. This model was first described by Sfetsos in \cite{Sfetsos:2014_integrability} as an interpolation between the exact CFT WZW model and the non-Abelian T-dual of the principal chiral model. It can be constructed by employing a particular gauging of the combined action for the gauged principal chiral model and the gauged WZW model. For the purpose of this work we are mostly interested in the final result of this procedure, and refer the interested reader to \cite{Sfetsos:2014_integrability,Hollowood:2014} for more details on the construction of the model itself. The action that describes the $\lambda$-deformed $G/G$ model is that of the $G/G$ model plus an additional deformation term and reads
\begin{equation}
	\label{eq:lambda_gWZW_action}
	\boxed{
		\quad S_{\lambda}[g,\mathbf{A}] = S_{G/G}[g,\mathbf{A}]+\frac{k}{\pi} \int d^2t\;\mathrm{Tr}\left(\gamma A\bar{A} \right)\,,\qquad \gamma = \lambda^{-1}-1\,,\quad}
\end{equation}
where the deformation is parametrized by $\gamma$ or $\lambda$. Clearly, for $\gamma=0$ (or, equivalently, $\lambda=1$) one recovers the ordinary $G/G$-model, while for $\gamma\rightarrow\infty$ (or, equivalently, $\lambda = 0)$ one recovers the WZW model when integrating out the gauge field.\footnote{To recover the non-Abelian T-dual of the principal chiral model, one should combine the limit $\lambda\rightarrow 1$ with the limit $k\rightarrow \infty$ in a particular way, see \cite{Sfetsos:2014_integrability}.} 

Some remarks are in order. To begin with, we stress that due to the deformation term in \eqref{eq:lambda_gWZW_action} the action $S_\lambda$ is no longer gauge invariant. As a result, the field $\mathbf{A}$ should no longer be interpreted as a gauge field, but rather as simply a constraint. However, we will still informally refer to $\mathbf{A}$ as the gauge field. Secondly, it should be noted that while the $\lambda$-deformed $G/G$ model is still conformal at the classical level, this is no longer true at the quantum level, in contrast to the $G/G$ model. This can be seen, for example, by analyzing the running of the coupling $\gamma$, whose $\beta$-function turns out to be non-trivial \cite{Tseytlin:1994,Sfetsos:2014_betafunction,Appadu:2015}. Nevertheless, the $\lambda$-deformed $G/G$ model retains some remarkable properties such as (strong) integrability \cite{Hollowood:2014,Sfetsos:2014_integrability,Georgiou:2019} and renormalizability \cite{Hoare:2019}. As a last remark, let us note that the action \eqref{eq:lambda_gWZW_action} also enjoys a discrete $\mathbb{Z}_2$ symmetry \cite{Kutasov:1989,Itsios:2014,Hoare:2015}
\begin{equation}\label{eq:Z2_symmetry}
	\lambda\mapsto \lambda^{-1}\,,\quad k\mapsto -k\,,\quad g\mapsto g^{-1}\,,\quad A\mapsto g(\partial + A)g^{-1}\,,\quad \bar{A}\mapsto \lambda^{-1}\bar{A}\,,
\end{equation}
which will be relevant in section \ref{sec:strings_Hodge_theory}. Physically, this symmetry can be viewed as a duality between a strong coupling regime $(\lambda\rightarrow \infty$) and a perturbative regime ($\lambda\rightarrow 0)$.

Let us now turn to the equations of motion of \eqref{eq:lambda_gWZW_action}, which are modified slightly compared to those of the $G/G$ model due to the deformation term. The equations of motion of $A$ and $\bar{A}$ are given by
\begin{align}
	\label{eq:eom_Abar_lambda}
	&\delta \bar{A}:\qquad g^{-1}Dg = \gamma A\,,\\
	\label{eq:eom_A_lambda}
	&\delta A:\qquad \bar{D}g g^{-1} = -\gamma \bar{A}\,,
\end{align}
while equation of motion of $g$ remains unchanged, but we record it here for convenience
\begin{equation}
	\label{eq:eom_g_lambda}
	\delta g:\qquad F-\bar{D}(g^{-1}Dg)=0\qquad \iff \qquad  F-D(\bar{D}g g^{-1}) = 0\,,
\end{equation}
where the equivalence of the two expressions can easily be verified.\footnote{This can be seen, for example, by writing the left-hand side as $[\partial+A+g^{-1}Dg, \bar{\partial}+\bar{A} ]=0$ and then conjugating this expression with $g$.} Our goal will be to simplify the above equations as much as possible. First, we note that \eqref{eq:eom_Abar_lambda} and \eqref{eq:eom_A_lambda} may be written as
\begin{align}
\label{eq:A_ginvdg}	&\delta\bar{A}:\qquad A = \frac{1}{\lambda^{-1}-\,\mathrm{Ad}_{g^{-1}}} g^{-1}\partial g\,,\\
\label{eq:Abar_ginvdbarg}	&\delta A:\qquad \bar{A} = \frac{1}{1-\lambda^{-1}\,\mathrm{Ad}_{g^{-1}}} g^{-1}\bar{\partial}g\,.
\end{align}
To be clear, the fraction in \eqref{eq:A_ginvdg} denotes the inverse of the linear operator $\lambda^{-1}-\mathrm{Ad}_{g^{-1}}$, and similarly in \eqref{eq:Abar_ginvdbarg}. While not immediately apparent, this form of the equations of motion will be extremely useful later on. 

Next, we consider the equation of motion of $g$. By inserting \eqref{eq:eom_Abar_lambda} into the left-hand side of \eqref{eq:eom_g_lambda} and \eqref{eq:eom_A_lambda} into the right-hand side of \eqref{eq:eom_g_lambda} one finds two equations
\begin{align}
	\label{eq:F_lambda}	&F-\gamma \bar{D}A = 0\qquad \iff \qquad  \lambda\partial\bar{A} - \bar{\partial}A+[A,\bar{A}] = 0\,,\\
	\label{eq:Fbar_lambda}	&F+\gamma D\bar{A}= 0\qquad \iff \qquad \partial\bar{A} - \lambda\bar{\partial}A+[A,\bar{A}] = 0\,.
\end{align}
When $\lambda=1$ (or $\gamma=0$, i.e.~no deformation) the equations \eqref{eq:F_lambda} and \eqref{eq:Fbar_lambda} coincide and one recovers the vanishing of the field-strength of the $G/G$ model. However, for $\lambda\neq 1$ (which we will henceforth assume) the field-strength need not vanish and the two equations are independent. By adding or subtracting the two equations appropriately one may rewrite them as follows
\begin{equation}\label{eq:eomg_full}
	\delta g:\qquad \partial\bar{A}+\bar{\partial}A = 0\,,\qquad \partial\bar{A} = \mu[A,\bar{A}]\,,\qquad \mu=-\frac{1}{1+\lambda}\,,
\end{equation}
where we also assume $\lambda\neq -1$. Passing to differential form notation, the equation on the left-hand side states that the one-form $\star\mathbf{A}$ is closed. In the remainder of this work, we will assume the worldsheet $\Sigma$ to be simply-connected by passing to the universal covering space. As a result, the one-form $\star\mathbf{A}$ is also exact, so that $\mathbf{A}$ must be of the form
\begin{equation}\label{eq:starA_closed}
	\mathbf{A} = \star\;\dd U\,,
\end{equation}
for some Lie-algebra valued function $U$. While not crucial, this observation will prove useful in section \ref{sec:strings_Hodge_theory}, where it will facilitate a natural ansatz for the gauge field.

We close this section by pointing out some possible generalizations of the (comparatively) simple $\lambda$-deformation we have considered that have been studied in the literature. One possibility is to consider $\lambda$-deformations of the $G/G$ model which do retain some of the gauge symmetry. These were first constructed for the $\mathrm{SU}(2)/\mathrm{U}(1)$ coset CFT in \cite{Sfetsos:2014_integrability}, and were later generalized to symmetric spaces $G/H$ in \cite{Hollowood:2014} and then applied to study the $AdS_5\times S^5$ superstring in \cite{Hollowood:2014qma}. Here the restriction to (semi)-symmetric spaces is crucial in order to retain properties such as (strong) integrability and renormalizability. Along a different vein, one may also consider deformations of multiple WZW models at different levels \cite{Georgiou:2018gpe}, as well as multi-parameter deformations \cite{Sfetsos:2015nya} and asymmetric deformations \cite{Georgiou:2016urf,Driezen:2019ykp}. For a recent overview of deformed $\sigma$-models that have arisen over the years and a discussion on their integrable structure, one may also consult e.g.~\cite{Georgiou:2021pbd}. Finally, in recent years great progress has been made on formulating integrable $\lambda$-deformations on worldsheets with boundaries and studying D-brane configurations, see e.g.~\cite{Alekseev:1998mc,Driezen:2018glg,Sfetsos:2021pcs}. 

\section{Variations of Hodge Structures}
\label{sec:Hodge_theory}

In this section we will describe a very different corner of physics/mathematics which, a priori, has very little to do with two-dimensional field theories, namely the study of variations of Hodge structures (VHS). We begin with a standard review on (polarized) Hodge structures in section \ref{subsec:HS} to set the notation and collect the various equivalent approaches that are used to describe a Hodge structure. In section \ref{subsec:VHS} we then describe a VHS, which can be seen as a collection of Hodge structures parametrized by a moduli space. We emphasize the formulation of a VHS in terms of the so-called period mapping and the charge operator. Of central importance is the horizontality condition \eqref{eq:horizontality_h} that the period mapping should satisfy, which greatly restricts its dependence on the moduli. It is this condition that will then be related to the equations of motion of the $\lambda$-deformed $G/G$ model in section \ref{sec:strings_Hodge_theory}. 

\subsection{Polarized Hodge Structures}
\label{subsec:HS}

Given a complex vector space $H$ and an integer $D$, a Hodge structure of weight $D$ can be described in a number of equivalent ways, all having their own merits.
The most common characterization of a Hodge structure is as a \textit{Hodge decomposition}, i.e.~a splitting of $H$ into $D+1$ pieces as follows
\begin{equation}
\label{eq:Hodge_decomposition}	H = H^{D,0}\oplus \cdots \oplus H^{0,D} = \bigoplus_{p+q=D} H^{p,q}\,,
\end{equation}
each satisfying $\overline{H^{p,q}} = H^{q,p}$ with respect to complex conjugation. We also introduce the notation $h^{p,q}=\mathrm{dim}\,H^{p,q}$. A familiar example of a Hodge decomposition is the splitting of the de Rham cohomology $H^D(X,\mathbb{C})$ of a K\"ahler manifold $X$ into the Dolbeault cohomology groups $H^{p,q}(X,\mathbb{C})$.

It will often be convenient to describe the decomposition \eqref{eq:Hodge_decomposition} in terms of an operator $Q$ that acts on $H$, which we will refer to as the \textit{charge operator}. More precisely, the $H^{p,q}$ are defined as the eigenspaces of $Q$ as follows
\begin{equation}
	Qv= (p-D/2)v\,,\qquad v\in H^{p,D-p}\,,\qquad p=0,\ldots,D
\end{equation}
such that $H^{p,D-p}$ is spanned by states of charge $p-D/2$. In particular, this implies that the adjoint action of $Q$ has an integer spectrum. Such an operator is also called a \textit{grading operator}. We will refer to the eigenstates of $Q$ as charge eigenstates, and to their eigenvalues as charges. In particular, the possible charges range from $D/2$ to $-D/2$, with eigenspaces corresponding to $H^{D,0}$ and $H^{0,D}$, respectively. Furthermore, the property $\overline{H^{p,q}} = H^{q,p}$ implies that $\overline{Q} = -Q$, so that $Q$ is a purely imaginary operator.

An equivalent formulation of a Hodge structure on $H$ is provided by a \textit{Hodge filtration}, i.e.~a decreasing filtration
\begin{equation}
	F^D\subseteq F^{D-1}\subseteq \cdots \subseteq F^0 = H\,,
\end{equation}
obtained by combining the different $(p,q)$-spaces in the following fashion
\begin{equation}
	F^p = H^{D,0}\oplus \cdots \oplus H^{D-p,p} = \bigoplus_{q\geq p} H^{q,D-q}\,,
\end{equation}
satisfying $H = F^p\oplus \overline{F^{D-p+1}}$. Indeed, one may recover the original $(p,q)$-spaces from the filtration via $H^{p,q}=F^p\cap \overline{F^q}$. In the language of the charge operator, each $F^p$ is spanned by states whose charge is greater or equal to $p-D/2$. Throughout the text, we will often switch between the description of the Hodge structure in terms of the decomposition, the charge operator and the filtration. 

Next, we impose an additional structure, namely that of a \textit{polarization}. This is a bilinear pairing on $H$, i.e.~a map
\begin{equation}
	(\cdot, \cdot): H\times H\rightarrow\mathbb{C}\,,
\end{equation}
which is required to be symmetric for $D$ even and skew-symmetric $D$ odd. Furthermore, it must satisfy the Hodge-Riemann bilinear relations 
\begin{align}
	&\text{(i)}:\qquad (H^{p,q}, H^{r,s}) = 0\,,\qquad \text{unless $p=s$ and $q=r$}\,,\\
	\label{eq:positivity}&\text{(ii)}:\qquad (\bar{v},Cv) > 0\,,\hspace{1.45cm} v\in H^{p,q}\,,v\neq 0\,.
\end{align}
Here the Weil operator $C$ is defined to act on $H^{p,q}$ as multiplication by $i^{p-q}$, which implies that it is related to the charge operator as $C=(-1)^{Q}$. The first relation simply states that the pairing of two charge eigenstates is zero unless their charges add to zero. Furthermore, by virtue of the second relation the pairing
\begin{equation}
	\langle v , w \rangle = (\bar{v}, Cw)
\end{equation}
defines an inner product, giving $H$ the structure of a Hilbert space. Finally, it will be useful to introduce $G$ as the group of real transformations that preserve the pairing $(\cdot,\cdot)$ and denote its algebra by $\mathfrak{g}$. Concretely, this means that
\begin{align}
\label{eq:def_G}	&g\in G:\qquad (gv, gw) = (v,w)\,,\\
\label{eq:def_g}	&X\in \mathfrak{g}:\qquad (Xv, w)+(v, Xw) = 0\,,
\end{align}
for all $v,w\in H$. Since $(\cdot,\cdot)$ is assumed to be either symmetric or skew-symmetric, one finds
\begin{equation}
	G = \begin{cases}
		\mathrm{Sp}(2n,\mathbb{R})\,, & \text{$D$ odd,}\qquad \\
		\mathrm{SO}(r,s)\,, & \text{$D$ even,}
	\end{cases}
\end{equation} 
where
\begin{equation}
	2m = \sum_{p} h^{p,D-p} = \mathrm{dim}\,H\,,\qquad r = \sum_{\text{$p$ even}} h^{p,D-p}\,,\quad s= \sum_{\text{$p$ odd}} h^{p,D-p}\,,\qquad r+s=2m\,.
\end{equation}
Similarly, we introduce $K$ as the group of real transformations that preserve the inner product $\langle\cdot,\cdot\rangle$ and denote its algebra by $\mathfrak{k}$. In other words, $K$ consists of unitary operators with respect to the given inner product. In particular, this implies that $K$ is a maximal compact subgroup of $G$ and is given by
\begin{equation}
	K = \begin{cases}
		\mathrm{U}(m)\,, & \text{$D$ odd,}\\
		\mathrm{SO}(2m)\cap\left(\mathrm{O}(r)\times \mathrm{O}(s)\right)\,, & \text{$D$ even.}
	\end{cases}
\end{equation}
Furthermore, from their definition it is clear that $C\in K$ and $Q\in i \mathfrak{k}$ (recall that $Q$ is purely imaginary). Lastly, let us note that the quotient $G/K$ of a semisimple Lie group $G$ by its maximal compact subgroup $K$ always yields a symmetric space. 

\subsection{Variations of Polarized Hodge Structures and the Period Mapping}\label{subsec:VHS}

Having discussed the notion of a single polarized Hodge structure, a natural generalization is to consider a family of polarized Hodge structures $H^{p,q}(t_1,\ldots, t_n)$ depending on a set of complex parameters $\{t_1,\ldots, t_n\}$. In this work we will restrict to $n=1$, meaning there is only a single complex parameter $t$ taking values in a worldsheet $\Sigma$. Important for us will be the fact that the dependence on $t$ is not arbitrary, but rather has to satisfy the following two conditions
\begin{align}
\label{eq:transversality}	\text{Griffiths transversality}:\qquad \frac{\partial}{\partial t} F^{p}&\subseteq F^{p-1}\,,\\
\label{eq:holomorphicity}	\text{holomorphicity}:\qquad \frac{\partial}{\partial \bar{t}} F^{p}&\subseteq F^{p}\,.
\end{align}
These two conditions are motivated by the Hodge decomposition of the primitive middle de Rham cohomology of a K\"ahler manifold, or more generally of algebraic families of algebraic varieties, for which they were shown to hold by Griffiths \cite{Griffiths:1968_I,Griffiths:1968_II}. Any family of polarized Hodge structures satisfying \eqref{eq:transversality} and \eqref{eq:holomorphicity} is referred to as a \textit{variation of polarized Hodge structures} \cite{Griffiths:1970}. It should be stressed that such variations of polarized Hodge structure need not have a geometric origin in the form of K\"ahler manifolds (or algebraic varieties) and can studied at an abstract level. 

While elegant, the conditions \eqref{eq:transversality} and \eqref{eq:holomorphicity} are difficult to implement practically, as they involve differential equations on vector spaces. To alleviate this issue, we proceed by rephrasing the notion of a VHS in two steps. First, given some reference Hodge structure $H^{p,q}_{\mathrm{ref}}$, it is easy to see that there always exist a coordinate-dependent operator $h(t,\bar{t})\in G$ such that\footnote{Note that $h$ is a real operator by virtue of the property $\overline{H^{p,q}}=H^{q,p}$.}
\begin{equation}
\label{eq:def_h}	H^{p,q}(t,\bar{t}) = h(t,\bar{t})\cdot H^{p,q}_{\mathrm{ref}}.
\end{equation}
In other words, the group $G$ acts transitively on the set of all polarized Hodge structures. In this way, all the coordinate dependence is captured by the field $h$. Note, however, that \eqref{eq:def_h} only defines $h$ up to local right multiplication by group elements that leave $H^{p,q}_{\mathrm{ref}}$ invariant. Concretely, we introduce the subgroup
\begin{equation}
	V = \{g\in G: g H^{p,q}_{\mathrm{ref}} = H^{p,q}_{\mathrm{ref}},\,\forall p,q\}\,,
\end{equation}
which is the stabilizer of all the $H^{p,q}_{\mathrm{ref}}$ spaces. Equivalently, $V$ is generated by operators which commute with the charge operator $Q_{\mathrm{ref}}$, which implies that $V$ is contained in $K$. Explicitly, one finds
\begin{equation}
	V = \begin{cases}
		\prod_{p\leq l} \mathrm{U}(h^{p,D-p})\,, & D=2l+1\,,\\
		\mathrm{SO}(h^{l,l})\times \prod_{p<l} \mathrm{U}(h^{p,D-p})\,,  & D=2l\,.
	\end{cases}
\end{equation}
The field $h$ should therefore be interpreted as a map 
\begin{equation}
	h: \Sigma\rightarrow G/V\,,
\end{equation}
which is known as the \textit{period mapping}.\footnote{Strictly speaking, when $\Sigma$ is not simply connected the field $h$ may undergo a non-trivial monodromy transformation under the monodromy group $\Gamma$. Then $h$ drops to a map into the double coset space $\Gamma\backslash G/V$ and this latter map is what is meant by the period mapping. This aspect will be central when discussing 
boundary conditions in \cite{Grimm_WZW_II}.} It assigns to each point in $\Sigma$ an element in the coset space $G/V$, which in turn provides a Hodge structure via \eqref{eq:def_h}. In other words, the space $G/V$ parametrizes all the possible polarized Hodge structures and is commonly referred to as the \textit{classifying space} or \textit{period domain}. Finally, let us remark that, in contrast to the quotient $G/K$, the classifying space $G/V$ is a reductive space, but generically not a symmetric space.

The conditions \eqref{eq:transversality} and \eqref{eq:holomorphicity} can now be written as differential equations for $h$ as follows
\begin{equation}\label{eq:horizontality_h}
	(h^{-1}\partial h)F_{\mathrm{ref}}^p \subseteq F_{\mathrm{ref}}^{p-1}\,,\qquad  (h^{-1}\bar{\partial} h)F_{\mathrm{ref}}^p \subseteq F_{\mathrm{ref}}^p\,.
\end{equation}
These equations remain difficult to work with, as they are still vector space equations. Therefore, as the second step we use the characterization of the Hodge structure in terms of eigenspaces of the reference charge operator $Q_{\mathrm{ref}}$. Recall that $F^p_{\mathrm{ref}}$ is spanned by states with charge greater or equal to $p-D/2$. Therefore, the first condition in \eqref{eq:horizontality_h} states that the operator $h^{-1}\partial h$ can lower the charge of a state by at most one, while $h^{-1}\bar{\partial}h$ may not lower the charge at all. To describe this more quantitatively, one introduces the charge-decomposition of an operator $\mathcal{O}\in \mathfrak{g}_{\mathbb{C}}$ as\footnote{Since $Q$ is not a real operator, this decomposition necessarily requires one to move to the complexification $\mathfrak{g}_\mathbb{C}$ of $\mathfrak{g}$.}
\begin{equation}
	\mathcal{O} = \sum_{q} \mathcal{O}_q\,,
\end{equation}
where the charge modes $\mathcal{O}_q$ are defined via
\begin{equation}\label{eq:q_mode}
	[Q_{\mathrm{ref}}, \mathcal{O}_q] = q\, \mathcal{O}_q\,,\qquad q=-D,\ldots, D\,.
\end{equation}
Clearly then, the action of $\mathcal{O}_q$ on a charge eigenstate raises its charge by $q$. For convenience, we also introduce the notation
\begin{equation}\label{eq:def_B}
	\mathbf{B}=h^{-1}\dd h\,,\qquad B = h^{-1}\partial h\,,\qquad \bar{B} = h^{-1}\bar{\partial}h\,,
\end{equation}
Then the statement that $B$ lowers the charge by at most one and $\bar{B}$ may not lower the charge at all translates to
\begin{equation}\label{eq:Qmodes_t}
	B_q = 0\,,\quad q< -1\,,\qquad \bar{B}_q = 0\,,\quad q< 0\,.
\end{equation} 
By taking the complex conjugate of \eqref{eq:Qmodes_t} and recalling that $\bar{Q}_{\mathrm{ref}} = -Q_{\mathrm{ref}}$, we arrive at a similar condition
\begin{equation}\label{eq:Qmodes_tbar}
	\bar{B}_q = 0\,,\quad q> 1\,,\qquad B_q = 0\,,\quad q> 0\, .
\end{equation} 
Finally, combining \eqref{eq:Qmodes_t} and \eqref{eq:Qmodes_tbar} one obtains
\begin{align}\label{eq:horizontality_B}
\boxed{\quad \text{horizontality condition}:\qquad 
	B_{-q} = 0 = 
	\bar{B}_q \,,\quad \text{unless $q=0$ or $q=1$}\,.\quad }
\end{align}
In summary, we have rephrased the condition \eqref{eq:horizontality_h} into the statement that only components of $B$ and $\bar{B}$ with a particular charge can be non-zero. We will refer to \eqref{eq:horizontality_B} as the \textit{horizontality condition}. 

At first sight, it appears that the non-zero components of $B$ and $\bar{B}$ remain unrestricted by the horizontality condition. This is, however, not the case. Indeed, note that $B$ and $\bar{B}$ satisfy the following no-curvature condition
\begin{equation}\label{eq:no_curvature_B}
	\partial\bar{B} - \bar{\partial}B+[B,\bar{B}]=0\,,
\end{equation}
as is apparent from their definition. Then by projecting \eqref{eq:no_curvature_B} onto its various charge components and using \eqref{eq:horizontality_B} one obtains a set of differential equations that constrain the various components of $B$ and $\bar{B}$ as follows
\begin{align}
\label{eq:charge_plus1}	&\text{charge $+1$}:\qquad \partial\bar{B}_{+1}+[B_0,\bar{B}_{+1}] = 0\,,\\
\label{eq:charge_0}	&\text{charge $0$}:\qquad \partial\bar{B}_0-\bar{\partial}B_0 +[B_{-1},\bar{B}_{+1}] = 0\,,\\
\label{eq:charge_min1}	&\text{charge $-1$}:\qquad \bar{\partial}B_{-1}+[\bar{B}_{0},B_{-1}] = 0\,.
\end{align}
It can be shown that in asymptotic regions, corresponding to $\mathrm{Im}\,t\rightarrow \infty$, the above equations reduce to Nahm's equations \cite{schmid}.\footnote{To see this, one should first employ the gauge-fixing condition $B_0=\bar{B}_0$ in order to fix the gauge invariance of $h$ under local right multiplications.} However, we will informally also refer to \eqref{eq:charge_plus1}-\eqref{eq:charge_min1} as Nahm's equations outside of the asymptotic regions. For the interested reader, we remark that systematic tools have been developed to study solutions to Nahm's equations in asymptotic regions of the field space in the seminal works of \cite{schmid} and Cattani, Kaplan and Schmid \cite{CKS}. This has lead to an algorithmic procedure via which an approximate solution to the period mapping can be obtained from a simple set of `boundary data', see also \cite{Grimm:2020cda} and \cite{Grimm:2021ikg} for recent perspectives in the physics literature and explicit examples of this procedure. In an upcoming continuation of this work a physical perspective on the boundary conditions will be developed further \cite{Grimm_WZW_II}.

Before closing this section, it is worthwhile to rephrase the description of a VHS yet again, in order to make contact with field content of the $\lambda$-deformed $G/G$ model. Recall that the Hodge structure $H^{p,q}_{\mathrm{ref}}$ can be encoded as the eigenspaces of the reference charge operator $Q_{\mathrm{ref}}$. Naturally, one can also encode the full VHS $H^{p,q}(t,\bar{t})$ via a coordinate-dependent charge operator $Q(t,\bar{t})$, which is naturally related to $Q_{\mathrm{ref}}$ via the period mapping as
\begin{equation}
	Q(t,\bar{t}) = h Q_{\mathrm{ref}} h^{-1}\,,
\end{equation}
as can be inferred from \eqref{eq:def_h}. We will refer to $Q(t,\bar{t})$ as the `bulk charge operator'. Of course, the coordinate-dependence of $Q(t,\bar{t})$ is restricted by the horizontality condition, which can be translated into the following condition\footnote{To derive this, one simply evaluates $\partial Q = \partial(h Q_{\mathrm{ref}}h^{-1}) = h[h^{-1}\partial h, Q_{\mathrm{ref}}]h^{-1}=h B_{-1} h^{-1}$ and then computes the commutator with $Q$, where in the last equality we have used the horizontality condition \eqref{eq:horizontality_B}.}
\begin{equation}\label{eq:horizontality_Q}
	\text{horizontality condition}:\qquad [Q,\partial Q] = -\partial Q\,,\qquad [Q,\bar{\partial}Q] = \bar{\partial}Q\,.
\end{equation}
The upshot of this is that one can formulate the notion of a VHS purely at the level of the algebra $\mathfrak{g}_{\mathbb{C}}$. Indeed, one concludes that whenever one has an operator $Q\in\mathfrak{g}_{\mathbb{C}}$ and a representation $H$ of $Q$ such that
\begin{enumerate}
	\item $Q$ is a grading operator, i.e.~$\mathrm{ad}_Q$ has an integer spectrum,
	\item $\bar{Q}=-Q$,
	\item $Q$ satisfies the horizontality condition \eqref{eq:horizontality_Q},
\end{enumerate}
then the eigenspaces of the representation of $Q$ define a variation of Hodge structures on $H$.

This concludes the discussion on variations of Hodge structures and the period mapping. In short, we have shown that the data of a variation of Hodge structure can be equivalently formulated in terms of the period mapping $h$, satisfying the condition \eqref{eq:horizontality_B}, which furthermore restricts the non-zero components charge components of $B$ and $\bar{B}$ with respect to $Q_{\mathrm{ref}}$ to satisfy \eqref{eq:charge_plus1}-\eqref{eq:charge_min1}. Alternatively, it can be captured by a purely imaginary grading operator $Q(t,\bar{t})$ that repects the condition \eqref{eq:horizontality_Q}. In the next section we will use both perspectives to show that the notion of a VHS can be realized as a solution to the equations of motion of the $\lambda$-deformed $G/G$ model by an appropriate identification of $Q(t,\bar{t})$ with the fields $g$ and $U$.

\section{Strings on Classifying Spaces}
\label{sec:strings_Hodge_theory}

In section \ref{sec:lambda_WZW} we have introduced the $\lambda$-deformed $G/G$ model, a particular $\sigma$-model for a group-valued field $g$ and gauge field $\mathbf{A}$. This model describes the classical propagation of a string on a group manifold, in the presence of particular background fields. It was then described, in section \ref{sec:Hodge_theory}, how the classifying space $G/V$ naturally parametrizes all polarized Hodge structures. Furthermore, the moduli dependence in a variation of Hodge structures was encoded in terms of the period mapping $h$ taking values in this space, together with the horizontality condition \eqref{eq:horizontality_B}. A natural step, then, is to consider the motion of the string \textit{on the classifying space}, as dictated by the equations of motion of the $\lambda$-deformed $G/G$ model, and ask whether this describes a variation of Hodge structures. The purpose of this section is to study precisely this question. Indeed, our main result is that for a suitable choice of gauge field and $\lambda$, the equations of motion of the string indeed describe a VHS. Conversely, this implies that any one-parameter VHS provides a solution to the $\lambda$-deformed $G/G$ model. Along the way we will also identify a more general set of solutions, which satisfy a generalized set of Nahm's equations.

We recall the properties of the bulk charge operator $Q(t,\bar{t})$, introduced at the end of section \ref{sec:Hodge_theory}. It is a grading element of a semisimple Lie algebra $\mathfrak{g}_\mathbb{C}$, meaning that it is a semisimple operator that acts on $\mathfrak{g}_{\mathbb{C}}$ via the adjoint action by integer eigenvalues, see e.g.~\cite{Kerr2017}. Furthermore, it must satisfy $Q\in i \mathfrak{g}$. Whenever such a $Q$ is available, one can consider the following ansatz for the fields of the $\lambda$-deformed $G/G$ model
\begin{equation}\label{eq:ansatz}
	g = e^{i\beta Q}\,,\qquad U = \alpha\, iQ\,,
\end{equation}
for some and non-zero parameters $\alpha,\beta\in\mathbb{C}$. The precise result of this section, then, is that for a suitable choice of $\alpha,\beta$, this provides a solution to the equations of motion of the $\lambda$-deformed $G/G$ model if and only if $Q$ satisfies the horizontality condition \eqref{eq:horizontality_Q}.  Note moreover that one can consider this ansatz without reference to Hodge theory as the above conditions merely pertain to the algebra $\mathfrak{g}_{\mathbb{C}}$, although there is a natural correspondence between a $Q_{\mathrm{ref}}$ with the above properties and a Hodge structure \cite{robles2012schubert}.

Of course, the ansatz \eqref{eq:ansatz} is motivated by the idea of considering strings on classifying spaces of Hodge structures. Indeed, we have argued that $Q$ itself contains all the information necessary to describe a VHS. Therefore the ansatz \eqref{eq:ansatz} is quite natural, as it is the most general ansatz one can make using only $Q$ as input. It should be mentioned that if one desires the fields $g$ and $U$ to be real (which is required for the action \eqref{eq:lambda_gWZW_action} to be real) one must restrict these parameters to $\alpha,\beta\in\mathbb{R}$. However, we will not make this restriction and leave the parameters arbitrary for now. In fact, it will turn out to be crucial to allow $\alpha$ to take complex values.

To interpret this ansatz from a physical perspective, we recall the passage of an action of the type we consider \eqref{eq:lambda_gWZW_action} to a more traditional $\sigma$-model action. One writes
\begin{equation}\label{eq:g_to_X}
	g^{-1}\dd g = T_A \tensor{e}{^A_\mu}\dd X^\mu\,,
\end{equation}
where the $T_A$ denote a set of generators of the Lie algebra $\mathfrak{g}$ and $\tensor{e}{^A_\mu}$ the left-invariant	 vielbein on $G$, which map the tangent bundle $TG$ to $\mathfrak{g}$. Finally, $X^\mu$ denote a set of local coordinates on $G$. One can then show that the action of the $\lambda$-deformed $G/G$ model, evaluated on the constraints \eqref{eq:A_ginvdg},\eqref{eq:Abar_ginvdbarg}, can be written as \cite{Driezen:2018glg}
\begin{equation}\label{eq:sigma_model}
	S_\lambda[g] = \frac{1}{4\pi\alpha'}\int_\Sigma \mathcal{G}_{\mu\nu}\,\dd X^\mu\wedge\star \,\dd X^\nu-\mathcal{B}_{\mu\nu} \,\dd X^\mu\wedge \dd X^\nu\,,
\end{equation}
with $\alpha'=2$ and
\begin{align}
	ds^2 &= k\,\tensor{e}{^A_\mu}\tensor{e}{^B_\nu} \mathrm{Tr}\left(T_A\cdot\left(\mathcal{O}_{g^{-1}}+\mathcal{O}_g-1\right)T_B\right)\,\dd X^\mu\otimes\dd X^\nu\,,\\
	\mathcal{B} &= \mathcal{B}_{\mathrm{WZW}}+k \,\tensor{e}{^A_\mu}\tensor{e}{^B_\nu}\mathrm{Tr}\left(T_A\cdot\left(\mathcal{O}_{g^{-1}}-\mathcal{O}_g\right)T_B\right)\dd X^\mu\wedge\dd X^\nu\,.
\end{align}
Here $\mathcal{B}_{\mathrm{WZW}}$ is such that locally $\dd \mathcal{B}_{\mathrm{WZW}} = -\frac{2k}{3}g^{-1}\dd g\wedge g^{-1}\dd g\wedge g^{-1}\dd g$ and we have used the short-hand $\mathcal{O}_g = (1-\lambda\mathrm{Ad}_g)^{-1}$. Note that $\mathcal{O}_g\rightarrow 1$ for $\lambda\rightarrow 0$, from which one readily sees that the above $\sigma$-model reduces to that of the ordinary WZW model in this limit. 

Clearly, equation \eqref{eq:sigma_model} describes the propagation of a string in a background metric $\mathcal{G}_{\mu\nu}$ and $\mathcal{B}_{\mu\nu}$ field, with the $X^\mu$ describing the embedding of the worldsheet. Moreover, inserting the ansatz \eqref{eq:ansatz} into \eqref{eq:g_to_X} one sees that the bulk charge operator $Q$ plays the role of the coordinates $X^\mu$. In other words, the ansatz \eqref{eq:ansatz} simply proposes that the moduli dependence of the Hodge structure is described by the embedding of the string worldsheet in the classifying space. However, in order for the motion of the string to correctly yield a variation of Hodge structures one must consider a particular value of $\lambda$, which amounts to a specific choice of the background fields. It is the purpose of the rest of this section to make this statement precise, by evaluating the equations of motion on the ansatz \eqref{eq:ansatz} and analysing the possible solutions. 

\subsection{The VHS Solution}
\label{sec:VHS_solution}

While we have formulated the ansatz \eqref{eq:ansatz} in terms of the bulk charge operator $Q$, it will be convenient to rephrase it in terms of the period mapping $h$ and the reference charge operator $Q_{\mathrm{ref}}$, by recalling that $Q=h Q_{\mathrm{ref}}h^{-1}$. First, for notational convenience, let us introduce the parameter
\begin{equation}
	z = e^{i\beta}\,,
\end{equation}
such that $g=z^{Q}$. Then one may compute
\begin{align}
	g^{-1}\dd g &= h z^{-Q_{\mathrm{ref}}}h^{-1} \dd h z^{Q_{\mathrm{ref}}}h^{-1}-\dd h h^{-1}\\
	&= h\left( z^{-\mathrm{ad}\,Q_{\mathrm{ref}}} \mathbf{B} - \mathbf{B}\right)h^{-1}\\
	&= \sum_{q\neq 0}(z^{-q}-1) h\mathbf{B}_q h^{-1}
\end{align}
where we recall $\mathbf{B}=h^{-1}\dd h$ and in the last line we have decomposed $\mathbf{B}$ into its charge-modes w.r.t. $Q_{\mathrm{ref}}$ as in \eqref{eq:q_mode}. Similarly, we compute
\begin{align}
	\mathbf{A} &= \star\,\dd (h z^{Q_{\mathrm{ref}}} h^{-1})\\
	&= -i\alpha\star h[Q_{\mathrm{ref}},h^{-1}\dd h]h^{-1}\\
	&=-i\alpha\sum_{q\neq 0} q \star h\mathbf{B}_q h^{-1}\,.
\end{align}
Together, these two identities provide us with the following dictionary
\begin{align}
\label{eq:g_to_B}	g^{-1}\dd g &= -\sum_{q\neq 0}(1-z^{-q})\, h\mathbf{B}_q h^{-1}\,,\\
\label{eq:A_to_B}	\mathbf{A} &= -i\alpha \sum_{q\neq 0} q\, \star h\mathbf{B}_{q}h^{-1}\,,
\end{align}
which will be very useful in evaluating the equations of motion. The formulation in terms of charge modes w.r.t. $Q_{\mathrm{ref}}$ is especially helpful combined with the observation that
\begin{equation}
\mathrm{Ad}_{g} = h z^{\mathrm{ad}\,Q_{\mathrm{ref}}}h^{-1},
\end{equation}
which acts as multiplication by $z^{q}$ on $h \mathbf{B}_q h^{-1}$.

The strategy in the remainder of this work is to use the dictionary \eqref{eq:g_to_B} and \eqref{eq:A_to_B} to translate the equations of motion of the $\lambda$-deformed $G/G$ model into statements about the charge-modes of $\mathbf{B}$. The goal, then, is to show that these statements imply the horizontality condition \eqref{eq:horizontality_B}, so that any solution to the equations of motion describes a variation of Hodge structures. However, it will turn out that this only happens for a particular value of $z$, namely $z=-1$ (or equivalently $\beta=\pi$). This is not unexpected, since in that case $g=(-1)^Q$, i.e.~it is precisely given by the Weil operator, previously denoted by $C$. Therefore, in the following we will first analyse this solution, dubbed the `VHS solution'. For completeness, we also present a more general analysis (i.e.~for all values of $z$) in the next section. 

\subsubsection{Equations of Motion of $A$ and $\bar{A}$}

For $z=-1$, equation \eqref{eq:g_to_B} simplifies to
\begin{equation}
	g^{-1}\dd g = -2\sum_{q\, \mathrm{ odd}} h \mathbf{B}_q h^{-1}\,,
\end{equation}
and furthermore $\mathrm{Ad}_g$ acts as multiplication by minus one on $\mathbf{B}_q$, for $q$ odd. With these observations at hand one can easily evaluate \eqref{eq:eom_Abar_lambda} and \eqref{eq:eom_A_lambda} to find
\begin{align}
\label{eq:eom_Abar_B}\delta\bar{A}:\qquad	\alpha \sum_{q\neq 0} q B_{q}= -\frac{2\lambda}{1+\lambda}\sum_{q\, \mathrm{ odd}}B_q\,,\\
\label{eq:eom_A_Bbar}\delta A:\qquad	\alpha \sum_{q\neq 0} q \bar{B}_{q}=+\frac{2\lambda}{1+\lambda}\sum_{q\, \mathrm{ odd}}\bar{B}_q\,.
\end{align}
These equations are best understood by projecting them onto a specific charge-mode, which we denote by $q$. Then one finds 
\begin{align}
	\text{$q$ non-zero and even}:\qquad &B_{q} = \bar{B}_{q}=0\,,\\
	\text{$q$ odd}:\qquad &b(q)\bar{B}_{q}=b(q)B_{-q}=0\,,
\end{align}
where we have introduced the function
\begin{equation}
	b(q)=\alpha q - \frac{2\lambda}{1+\lambda}\,.
\end{equation}
The upshot of this is the following. First, $\mathbf{B}$ cannot contain any non-zero even charge-modes. Second, it can only contain an odd charge-mode $q$ if $b(q)=0$. Furthermore, if $b(q_i)=0$ for two distinct odd $q_i$, $i=1,2$, then it must be that $\alpha=\lambda=0$. In that case the ansatz for the gauge field becomes trivial, hence we exclude this solution. Therefore $b(q)=0$ for at most one odd $q$. In short, we have shown that $\mathbf{B}$ must be of the form
\begin{equation}\label{eq:B_simplified}
	B = B_0 + B_{-q}\,,\qquad \bar{B}=\bar{B}_0+\bar{B}_q\,,
\end{equation}
for some odd $q$, and furthermore the parameters $\alpha,\lambda$ are restricted by
\begin{equation}\label{eq:sol_alpha}
	\alpha q = \frac{2\lambda}{1+\lambda}\,,
\end{equation}
so that indeed $b(q)=0$. This is then the most general solution to the equations of motion of $A$ and $\bar{A}$, for the ansatz \eqref{eq:ansatz} when imposing $\beta=\pi$. As promised, \eqref{eq:B_simplified} is precisely the desired horizontality condition \eqref{eq:horizontality_B}. To make the match precise, one should also fix $\alpha$ to be $2\lambda/(1+\lambda)$, which effectively enforces $q=1$. 

\subsubsection{Equation of motion of $g$}

While the equations of motion of $A$ and $\bar{A}$ restrict the possible charge-modes in $B$ and $\bar{B}$, as just discussed, the equation of motion of $B$ will determine the remaining dynamics and fix the value of $\lambda$ uniquely. Of course, as already described in section \ref{sec:Hodge_theory}, once the horizontality condition is satisfied one can use the Bianchi identity to determine the dynamics completely, which resulted in Nahm's equations. Here we will show that the equations of motion of $g$ essentially reduce to the same equations.

First, using \eqref{eq:B_simplified}, one sees that the gauge field simplifies to
\begin{equation}
	A = -\alpha q\, hB_{-q}h^{-1}\,,\qquad \bar{A} = -\alpha q\, h \bar{B}_{q}h^{-1}\,.
\end{equation}
Inserting these expressions into the equation of motion of $g$ \eqref{eq:eomg_full} yields the following equation
\begin{equation}
	\delta g:\qquad \partial\bar{B}_{q}+[B_0,\bar{B}_q]+(1+q\alpha\mu)[B_{-q},\bar{B}_q]=0\,.
\end{equation}
Again, this equation is best understood by considering its various charge components. In particular, the first two terms have charge $q$, while the last term has charge zero. In other words, it reduces to the following two equations
\begin{align}
	&\text{charge $q$}:\qquad \partial\bar{B}_{q}+[B_0,\bar{B}_q]=0\,,\\
	&\text{charge 0}:\qquad (1+q\alpha\mu)[B_{-q},\bar{B}_q]=0\,.
\end{align}
Naturally, by taking the complex conjugate of the first equation, one may equivalently derive
\begin{equation}
	\text{charge $-q$}:\qquad \bar{\partial}B_{-q}+[\bar{B}_0,B_{-q}]=0\,.
\end{equation}
As expected, the equation of motion of $g$ has reduced to the same equations that follow from the Bianchi identity, i.e.~Nahm's equations \eqref{eq:charge_plus1} and \eqref{eq:charge_min1}, together with an additional constraint on the parameters, namely $1+q\alpha \mu=0$. Combined with the constraint \eqref{eq:sol_alpha} this results in a unique solution (up to a sign) for $\alpha$ and $\lambda$, given by
\begin{equation}\label{eq:alpha_sol}
	q\alpha = 1+\lambda\,,\qquad \lambda = \pm i.
\end{equation} 
Furthermore, one can show that these two solutions are precisely related via the $\mathbb{Z}_2$-symmetry \eqref{eq:Z2_symmetry}. 

To close this discussion, we stress that we have not actually solved the equations of motion in full, but have merely rewritten them into the horizontality condition and Nahm's equations. However, as described mentioned in section \ref{subsec:VHS}, the powerful formalism of asymptotic Hodge theory allows one to find the proper solutions to these equations for a particular set of boundary conditions.

\subsubsection{Integrability and the On-Shell Action}

Now that we have identified for which ansatz the equations of motion of the $\lambda$-deformed $G/G$ model reduce to the horizontality condition, let us return to some properties of the $\sigma$-model. Firstly, as mentioned before, this model is integrable, i.e.~it has a Lax connection 
\begin{equation}
	\nabla = \dd +\mathcal{L}(\zeta)\,,
\end{equation}
with $\mathcal{L}(\zeta)$ given by (for notational clarity we use $\pm$ to denote the (anti)-holomorphic components of the one-form $\mathcal{L}$ and $\mathbf{A}$)
\begin{equation}\label{eq:Lax}
	\mathcal{L}_\pm(\zeta)=\frac{2}{1\mp \zeta} \frac{A_{\pm}}{1+\lambda}\,.
\end{equation}
Concretely, this means that the curvature of the connection $\nabla$ vanishes if and only if $A_\pm$ satisfy \eqref{eq:F_lambda} and \eqref{eq:Fbar_lambda}, for every value of the spectral parameter $\zeta\in\mathbb{C}\mathbb{P}^1$. It is natural, then, to evaluate \eqref{eq:Lax} on the VHS solution to find an expression for $\mathcal{L}(\zeta)$ in terms of $B$ and $\bar{B}$. The result of this computation is the following
\begin{equation}
	\mathcal{L}_\pm(\zeta) = -\frac{2}{1\mp\zeta} h \left(B_{\pm}\right)_{\pm q}h^{-1}\,,
\end{equation}
where we have used \eqref{eq:alpha_sol}, for arbitrary $q$. To elucidate this expression, we perform a gauge transformation by the period mapping itself, i.e.
\begin{equation}
	\mathcal{L}_\pm(\zeta)\rightarrow \mathcal{L}_{\pm}^{h^{-1}}(\zeta) = h^{-1}\mathcal{L}_{\pm}(\zeta)h + h^{-1}\partial_\pm h\,,
\end{equation} 
which yields
\begin{equation}\label{eq:Lax_gauge_transform}
	\mathcal{L}^{h^{-1}}_\pm(\zeta) = (B_\pm)_0 + \frac{\zeta\pm 1}{\zeta\mp 1} (B_{\pm})_{\pm q}\,.
\end{equation}
Interestingly, \eqref{eq:Lax_gauge_transform} is precisely the Lax connection of the $G/V$ principal chiral model \cite{Hollowood:2014}, where we remind the reader that in the current setting $V$ is generated by all operators of zero charge. In other words, when the horizontality condition is satisfied, the remaining dynamics of the period mapping are described by the integrable $G/V$ principal chiral model. This is not surprising and has long been known from the perspective of $tt^*$-geometry \cite{Cecotti:1991me,Cecotti:2020rjq} and was more recently noted by \cite{Cecotti:2020uek} and \cite{Grimm:2020cda}.

To further elucidate the relation to these previous works, it is instructive to compute the on-shell action (i.e.~by imposing \eqref{eq:A_ginvdg} and \eqref{eq:Abar_ginvdbarg}) and rephrasing the result in terms of $B$ and $\bar{B}$. The result of this computation is (see appendix \ref{sec:app_action} for details)
\begin{equation}\label{eq:on-shell_action}
	S_\lambda[g] = \pm \frac{2k}{\pi i}\int_\Sigma d^2t \,\mathrm{Tr}\left(B_{-1}\bar{B}_{+1}\right)\,,
\end{equation}
where we have put $q=1$ for simplicity. Apart from the imaginary prefactor, this agrees precisely with the actions put forward in \cite{Cecotti:2020uek} and \cite{Grimm:2020cda,Grimm:2021ikg}. Therefore, the formulation in terms of the $\lambda$-deformed $G/G$ model can be seen as a generalization of these earlier works, which reduces to them when $A$ and $\bar{A}$ are on-shell, i.e.~when the horizontality condition is satisfied. 

As a final remark, we comment on the imaginary prefactor in \eqref{eq:on-shell_action}, which is a result of the fact that both $\alpha$ and $\lambda$ are complex. Interestingly, a similar situation arises in the study of a duality between $\lambda$-deformations and so-called $\eta$-deformations \cite{Klimcik_2015}, where we remark that also in our setting $|\lambda|=1$, as is required for this duality to apply. In fact, a more general analysis of the equations of motion shows that whenever $|z|=1$ also $|\lambda|=1$, as is discussed in the next section. It would be interesting to understand this duality from the perspective of variations of Hodge structures, which may provide new insights into the structure of classifying spaces.

\subsection{General Analysis of the Equations of Motion}

In the following, we relax the assumption that $z=-1$ and go through the same analysis as before, which now becomes considerably more involved. In particular, it becomes possible for $B$ and $\bar{B}$ to have more than one charge-mode, hence this solution will not satisfy the horizontality condition. Nevertheless, it turns out that this solution is still rather constrained, with its various charge-modes independently satisfying a variant of Nahm's equations. 

\subsubsection*{Equations of Motion of $A$ and $\bar{A}$}

We recall the relations \eqref{eq:g_to_B} and \eqref{eq:A_to_B} and insert them into the equations of motion of $A$ and $\bar{A}$, see \eqref{eq:eom_Abar_lambda} and \eqref{eq:eom_A_lambda}. Then one finds
\begin{align}
	&\delta\bar{A}:\qquad \alpha \sum_{q\neq 0} q B_{q}= -\sum_{q\neq 0}\frac{1-z^{-q}}{\lambda^{-1}-z^{-q}}B_q\,,\\
	&\delta A:\qquad\alpha \sum_{q\neq 0} q \bar{B}_{q}=+\sum_{q\neq 0}\frac{1-z^{-q}}{1-\lambda^{-1}z^{-q}}\bar{B}_q\,.
\end{align}
Note that for $z=-1$ this indeed reduces to \eqref{eq:eom_Abar_B} and \eqref{eq:eom_A_Bbar}. Since both equations should hold for each $q$ separately, this yields the following constraints
\begin{align}\label{eq:beta}
	b(q)B_{-q} & =0\,, \qquad b(q)\bar{B}_q = 0\,,\qquad q\neq 0\,,
\end{align}
where we have defined the function
\begin{equation}\label{eq:def_beta}
	b(q)=\alpha q - \frac{1-z^{q}}{\lambda^{-1}-z^{q}}\,.
\end{equation}
By a slight abuse of notation we use the same symbol $b(q)$ as in section \ref{sec:VHS_solution}, but of course the two agree for $z=-1$. However, the important observation is that for $z\neq -1$, one can actually solve $b(q)=0$ simultaneously for multiple values of $q$. 

Let us therefore denote by $\mathbf{q}=(q_1,\ldots, q_n)$ a vector of distinct non-zero charges $q_i$ for which $b(q_i)=0$. Clearly, \eqref{eq:beta} implies that $\bar{B}_q=B_{-q}=0$ for all $q\not\in\mathbf{q}$. Therefore, to understand which combinations of charge-modes in $B$ and $\bar{B}$ are allowed, one should study the zeroes of $b(q)$. At this point, it is convenient to distinguish the trivial solutions to $b(q)=0$, corresponding to $\alpha=0$ and $\lambda=1$, or $\alpha=0$ and $z^{q}=1$. Furthermore, as alluded to in section \ref{sec:lambda_WZW}, we also assume $\lambda\neq -1$ and also refer to this as a trivial solution. In the following, whenever we speak of a solution, it is implicitly implied to be non-trivial.

Below one finds a number of properties of $b(q)$ that will be useful in the remainder of this work. The first property is obvious, whereas the latter properties are elaborated upon in appendix \ref{sec:app_beta}. 
\begin{align}
	&\text{property (1)}:\qquad b(q) = b(-q)=0\quad\implies\quad \lambda = -1\,.\\
\nonumber	&\text{property (2)}:\qquad \text{For $q_1,q_2$ distinct, with $b(q_1)=b(q_2)=0$, the following holds:}\\
	&\hspace{3cm} \text{(2a)}:\qquad b(2q_2)=0\quad\hspace{0.65cm}\iff\quad b(2q_2-q_1)=0\,,\\
\label{eq:2b}	&\hspace{3cm}\text{(2b)}:\qquad b(q_1-q_2)=0\quad\iff\quad 1+q_1\alpha \mu=0\,\\
\label{eq:2c}	&\hspace{3cm}\text{(2c)}:\qquad b(q_1+q_2)=0\quad\hspace{0.05cm}\iff\quad 1+(q_1+q_2)\alpha\mu=0\,.
\end{align}
Note that property (2a) can be derived from (2b) and (2c), but we still write it down separately for easy reference later on. Property (1) implies that it is impossible for both $b(q)$ and $b(-q)$ to be zero, when restricting to non-trivial solutions. Therefore, it is impossible to have $q\in\mathbf{q}$ and $-q\in\mathbf{q}$. One must therefore take $B$ and $\bar{B}$ to be of the form
\begin{equation}\label{eq:B_form}
	B = B_0 + B_{-q_1}+\cdots B_{-q_n}\,,\qquad \bar{B}=\bar{B}_0 + \bar{B}_{q_1}+\cdots+\bar{B}_{q_n}\,,
\end{equation}
where $q_i\neq 0$ and $q_i\neq \pm q_j$ for any distinct $i,j$. Furthermore, we remind the reader that all these charges should satisfy $b(q_i)=0$. We stress that this solves the equations of motion of $A$ and $\bar{A}$, and that all solutions respecting the ansatz \eqref{eq:ansatz} must be of this form. 

A natural question is whether there is an upper bound on $n$, i.e.~the number of non-zero charge components of $B$ and $\bar{B}$, above which no solutions exist. Intuitively, this is expected since $b(q_1)=\ldots=b(q_n)=0$ constitutes $n$ complex equations, whereas there are only three free complex parameters $\alpha,\lambda, z$. Naively, one therefore expects $n\leq 3$. However, property (2a) implies that the system of equations can degenerate when $q_3=2q_2$ and $q_4=2q_2-q_1$. Hence solutions with $n=4$ do exist. In fact, a numerical scan\footnote{This constitutes 11176 potential charge vectors $\mathbf{q}$, of which 57 yield non-trivial solutions.} indicates that for $D\leq 15$ all solutions with $n=4$ are of this form. Furthermore, another numerical scan\footnote{This constitutes 48913 potential charge vectors $\mathbf{q}$, of which 0 yield non-trivial solutions.} shows that for $D\leq 15$, no solutions with $n=5$ exist. In summary, we have made the following two claims
\begin{align}
	&\text{claim (1)}:\qquad \text{For $n=4$, all solutions have $\mathbf{q}=(q_1, q_2, 2q_2, 2q_2-q_1)$}\,,\\
	&\text{claim (2)}:\qquad \text{No solutions exist with $n>4$.}
\end{align}
In the remainder of this work, we will assume that these two claims are true for all $D$. 

\subsubsection*{Equation of Motion of $g$}

Let us now discuss the remaining equation of motion, which we recall for convenience
\begin{equation}
	\delta g:\qquad \partial\bar{A} = \mu[A,\bar{A}]\,,\qquad \mu = -\frac{1}{1+\lambda}\,.
\end{equation}
Again, our goal is to evaluate this equation on the ansatz \eqref{eq:ansatz} to translate it into a condition on the various charge modes of $\mathbf{B}$. In appendix \ref{sec:app_eomg} this computation is performed in detail, here we simply record the result (see equations \eqref{eq:Bbar} and \eqref{eq:B})
\begin{align}
\label{eq:eomg_qstar_1}	0&=q^*\left(\partial\bar{B}_{q^*}+[B_0,\bar{B}_{q^*}]\right) + \sum_{q\neq 0,q^*}q (1+(q-q^*)\alpha\mu)[B_{q^*-q}, \bar{B}_{q}]\,,\\
\label{eq:eomg_qstar_2}	0&=q^*\left(\bar{\partial}B_{q^*}+[\bar{B}_0,B_{q^*}]\right)+\sum_{q\neq 0,q^*}(q-q^*)(1+q\alpha\mu)[B_{q^*-q}, \bar{B}_{q}]\,.
\end{align}
To be clear, here we have projected the resulting equations onto a general charge-mode, denoted by $q^*$. In principle, these two equations are not independent as one can derive one from the other by using the flatness of $\mathbf{B}$, i.e.~the fact that $\partial\bar{B}-\bar{\partial}B+[B,\bar{B}]=0$. However, in practice it will be useful to use both equations as opposed to the flatness condition. One sees that the equation of motion of $g$ has reduced to a set of differential equations for the various charge components of $B$ and $\bar{B}$ that is similar to Nahm's equations \eqref{eq:charge_plus1}-\eqref{eq:charge_min1}, but is spoiled by the additional cross-terms appearing in the sums. We will now argue that these additional terms take a very simple form. To do so, it is convenient to distinguish the four types of charge-modes that can appear in \eqref{eq:eomg_qstar_1} and \eqref{eq:eomg_qstar_2}.

\subsubsection*{Type (1): $q^*=0$}

When $q^*=0$, the equations \eqref{eq:eomg_qstar_1} and \eqref{eq:eomg_qstar_2} simply reduce to a single equation, namely
\begin{equation}
	0 =\sum_{q\neq 0} q(1+q\alpha\mu)[B_{-q},\bar{B}_{q}]\,.
\end{equation}
Note that it is impossible for the prefactor $1+q\alpha\mu$ to vanish for more than one value of $q$, so this yields a proper constraint on the commutators of the form $[B_{-q},\bar{B}_q]$. 

\subsubsection*{Type (2): $q^*\in\mathbf{q}$}

Suppose there exists a $q\in\mathbf{q}$ such that $\bar{B}_q\neq 0$ and also $B_{q^*-q}\neq 0$. In particular, this requires $q,q-q^*\in\mathbf{q}$. By property (2b), it follows that $1+q\alpha\mu=0$. Therefore, if such a $q$ exists it must be unique. In fact, there can be at most one $q^*\in\mathbf{q}$ for which this is the case. Finally, since $q^*\in\mathbf{q}$ it is impossible to have $-q^*\in\mathbf{q}$, hence it must be that $B_{q^*}=0$. As a result, equation \eqref{eq:eomg_qstar_2} is automatically satisfied, while \eqref{eq:eomg_qstar_1} reduces to
\begin{equation}\label{eq:eomg_case2}
	\partial\bar{B}_{q^*}+[B_0,\bar{B}_{q^*}]+[B_{q^*-q},\bar{B}_q]=0\,.
\end{equation}
Of course, if a $q$ with the above properties does not exist, the last term in this expression is simply absent. 

\subsubsection*{Type (3): $-q^*\in\mathbf{q}$}

As in case (2), suppose there exists a $q\in\mathbf{q}$ such that $\bar{B}_q\neq 0$ and also $B_{q^*-q}\neq 0$, which implies $q,q-q^*\in\mathbf{q}$. By property (2c), it follows that $1+(q-q^*)\alpha\mu=0$, hence the sum in \eqref{eq:eomg_qstar_1} is zero. Furthermore, since already $-q^*\in\mathbf{q}$ it is impossible to have $q^*\in\mathbf{q}$, hence $\bar{B}_{q^*}=0$. In other words, in this case equation \eqref{eq:eomg_qstar_1} is automatically satisfied. Moreover, equation \eqref{eq:eomg_qstar_2} reduces to
\begin{equation}
	\bar{\partial}B_{q^*}+[\bar{B}_0,B_{q^*}]-[B_{q^*-q},\bar{B}_q]=0\,.
\end{equation}
Note that this is simply the complex conjugate of \eqref{eq:eomg_case2} with $q^*\mapsto -q^*$ and $q\mapsto q-q^*$, as expected. 

\subsubsection*{Type (4): $\pm q^*\not\in\mathbf{q}$ and $q^*\neq 0$}

Finally, we discuss the remaining case, which will turn out to be most complex. First, since $\pm q^*\not\in\mathbf{q}$ and $q^*\neq 0$, it follows that $\bar{B}_{q^*}=0$ and $B_{q^*}=0$, so \eqref{eq:eomg_qstar_1} and \eqref{eq:eomg_qstar_2} reduce to 
\begin{align}
	\label{eq:eomg_qstar_1_case4}	0&= \sum_{q\neq 0,q^*}q (1+(q-q^*)\alpha\mu)[B_{q^*-q}, \bar{B}_{q}]\,,\\
	\label{eq:eomg_qstar_2_case4}	0&=\sum_{q\neq 0,q^*}(q-q^*)(1+q\alpha\mu)[B_{q^*-q}, \bar{B}_{q}]\,.
\end{align}
To continue, it will be useful to introduce the `multiplicity' $m(q^*)$ of $q^*$, by which we mean the number of distinct pairs $q_i,q_j\in\mathbf{q}$ whose difference is equal to $q^*$. In other words, $m(q^*)$ counts the number of terms appearing in the sums in \eqref{eq:eomg_qstar_1_case4} and \eqref{eq:eomg_qstar_2_case4}. It is clear that for $n\leq 3$, one can have at most multiplicity two. Furthermore, if $n=4$ one may apply claim (1) to see that the only way to have $m(q^*)=3$ is for $\mathbf{q}$ to be of the form $\mathbf{q}=(2q,q,4q,3q)$. However, one can check that such a $\mathbf{q}$ cannot yield non-trivial solutions to $b(\mathbf{q})=0$. Therefore, since we assume $n\leq 4$ following claim (2), we may restrict to the cases $m(q^*)=1,2$. These will now be discussed in turn. 

\subsubsection*{Type (4a): $m(q^*)=1$}

Let $q_i,q_j\in\mathbf{q}$ and suppose that the difference $q^*=q_i-q_j$ has multiplicity one. In the following, we again assume $\bar{B}_q\neq 0$ for $q=q_i,q_j$, otherwise \eqref{eq:eomg_qstar_1_case4} and \eqref{eq:eomg_qstar_2_case4} are trivially satisfied. Since $m(q^*)=1$, they simply reduce to
\begin{align}
	(1+q_j\alpha \mu)[B_{-q_j},\bar{B}_{q_i}]&=0\,,\\
	(1+q_i\alpha\mu)[B_{-q_j},\bar{B}_{q_i}]&=0\,.
\end{align}
Clearly, since $q_i\neq q_j$ this can hold only when $[B_{-q_j},\bar{B}_{q_i}]=0$. 

\subsubsection*{Type (4b): $m(q^*)=2$}

Next, let $q_i,q_j,q_k,q_l\in\mathbf{q}$ such that the difference $q^*=q_i-q_j=q_k-q_l$ has multiplicity two, which requires $q_i\neq q_k$ and $q_j\neq q_l$. In the following, we again assume $\bar{B}_{q}\neq 0$ for $q=q_i,q_j,q_k,q_l$, otherwise \eqref{eq:eomg_qstar_1_case4} and \eqref{eq:eomg_qstar_2_case4} are trivially satisfied. We turn to \eqref{eq:eomg_qstar_1_case4} and \eqref{eq:eomg_qstar_2_case4}, which reduce to
\begin{align}
	\text{$\bar{B}_{q^*} = 0$}:\qquad q_i(1+q_j\alpha \mu)[B_{-q_j},\bar{B}_{q_i}]+ q_k(1+q_l\alpha \mu)[B_{-q_l},\bar{B}_{q_k}]&=0\,,\\
	\text{$B_{q^*}=0$}:\qquad q_j(1+q_i\alpha\mu)[B_{-q_j},\bar{B}_{q_i}]+q_l(1+q_k\alpha\mu)[B_{-q_l},\bar{B}_{q_k}]&=0\,.
\end{align}
Since both equations must hold, one may combine them succinctly as a matrix equation
\begin{equation}\label{eq:eomg_M}
	M\cdot \begin{pmatrix}
		[B_{-q_j},\bar{B}_{q_i}]\\ [B_{-q_l},\bar{B}_{q_k}] 
	\end{pmatrix}=0\,,\qquad M = \begin{pmatrix}
		q_i(1+q_j\alpha\mu) & q_k(1+q_l\alpha\mu)\\
		q_j(1+q_i\alpha\mu) & q_l(1+q_k\alpha\mu)
	\end{pmatrix}\,,
\end{equation}
with
\begin{align}
	\det M &= (q_i q_l - q_j q_k) - \left[q_i q_j q_k+q_j q_k q_l - q_i q_j q_l - q_i q_k q_l\right]\alpha\mu\,.
\end{align}
Using the fact that $q_i-q_j=q_k-q_l$, one finds that
\begin{equation}
	\det M = 0\quad\iff\quad 1+(q_j+q_k)\alpha\mu = 0\,.
\end{equation}
From property (2c) it then follows that $\det M=0$ if and only if $q_j+q_k\in\mathbf{q}$. In particular, the equation \eqref{eq:eomg_M} only has a non-trivial solution when $q_j+q_k\in\mathbf{q}$. In summary, we have shown that
\begin{equation}
		[B_{-q_j},\bar{B}_{q_i}] = [B_{-q_l},\bar{B}_{q_k}]=0\quad\text{unless}\quad q_j+q_k\in\mathbf{q}\,.
\end{equation}
It is interesting to note that this precisely occurs in the case $n=4$, for which the charge vectors take the form (recall claim (1))
\begin{equation}
	\mathbf{q} = (q_1, q_2, 2q_2, 2q_2-q_1)\,.
\end{equation}
Indeed, one has $(2q_2-q_1)-(q_2) = q_2-q_1$ and additionally $2q_2\in\mathbf{q}$, so that a non-trivial solution exists. 

\subsubsection*{Summary of the Analysis}

For the convenience of the reader, let us collect the results of this section. To simplify the overview, we only consider the $\bar{B}$ component. Statements about the $B$ component are naturally obtained by complex conjugation. First, by studying the equations of motion of $A$ and $\bar{A}$, it was argued that the charge decomposition of $\bar{B}$ must take the following form
\begin{equation}\label{eq:sol_barB}
	\bar{B} = \bar{B}_0+\bar{B}_{q_1}+\cdots+\bar{B}_{q_n}\,,
\end{equation}
where $q_i\neq 0$ and furthermore $q_i\neq\pm q_j$ for any $i\neq j$. In order to satisfy the equations of motion, the integers $q_i$ and parameters $\alpha,\lambda,z$ must be such that $b(q_i)=0$ for all $i$. Also, due to the particular structure of $b(q)$, it appears impossible to have $n>4$ based on numerical analysis, while for $n=4$ only special combinations of charges are allowed (recall claim (1)), although we have not rigorously proven this. 

Second, the dynamics of the $\bar{B}_{q_i}$ modes are restricted via the equation of motion of $g$. Concretely, we have found that they must satisfy
\begin{equation}\label{eq:dynamics_B}
	\partial\bar{B}_{q_i}+[B_0,\bar{B}_{q_i}]+[B_{q_i-q_j},\bar{B}_{q_j}]=0\,.
\end{equation}
Additionally, there exists at most one value of $q_i$ for which there exists a $q_j$ such that the last term is non-zero. In that case, also $q_j$ is unique. Strikingly, this means that all the $\bar{B}_{q_i}$ modes satisfy a variant of Nahm's equations \eqref{eq:charge_plus1}-\eqref{eq:charge_min1} apart from possibly one of the modes, for which this one additional term appears. 

Lastly, from the discussion of type (4), all the commutators of the form $[B_{-q_j},\bar{B}_{q_i}]$ with $q_i\neq q_j$ and $q_i-q_j\not\in\mathbf{q}$ must in fact vanish, except when $q_i-q_j=q_k-q_l$, for $q_i\neq q_k$, and additionally $q_j+q_k\in\mathbf{q}$. We stress that this conclusion only holds for $n\leq 4$. Moreover, when $q_i=q_j$ corresponding to type (1), then the commutators are restricted to satisfy
\begin{equation}\label{eq:constraint_commutator}
	\sum_{i=1}^n q_i(1+q_i\alpha\mu)[B_{-q_i},\bar{B}_{q_i}]=0\,.
\end{equation}
This concludes the analysis of the equations of motion. 

\subsubsection{Properties of Solutions}

We close this section with some final remarks on properties of solutions to the equations of motion, which shed further light on the properties we encountered for the VHS solution. To this end, it is convenient to distinguish the cases $n=1$ and $n\geq 1$, with $n$ counting the number of non-zero charge-modes of $B$ and $\bar{B}$, as in \eqref{eq:sol_barB}.

\subsubsection*{Case (I): $n=1$}

In this case one simply has $\bar{B} = \bar{B}_0 + \bar{B}_q$, as for the VHS solution, except for generic $z$ the condition $b(q)=0$ also allows for even values of $q$. In fact, one can solve the conditions $b(q)=0$ and $1+q\alpha\mu=0$ algebraically to find
\begin{equation}
	\alpha q = 1+\lambda\,\qquad \lambda = \pm z^{-q/2}\,.
\end{equation}
Note that for $z=-1$ and $q$ odd this precisely reduces to the VHS solution, as expected. Again the two solutions we find are related via the $\mathbb{Z}_2$ symmetry \eqref{eq:Z2_symmetry}. The main observation we would like to make is that whenever $|z|=1$, which was required for $g$ to be real, it must again be that also $|\lambda| =1$.

\subsubsection*{Case (II): $n\geq 2$}

When $n\geq 2$, one needs to solve $b(q_1)=\cdots=b(q_n)=0$. It is actually possible to solve $b(q_1)=b(q_2)=0$ algebraically for $\alpha$ and $\lambda$, with the result given by
\begin{align}
	b(q_1)=b(q_2)=0\quad\iff\quad \alpha = -\frac{F(z^{q_1},z^{q_2})}{F(q_2,q_1)}\,,\qquad \lambda  = \frac{F(z^{q_1},z^{q_2})}{F(1,1)}\,,
\end{align}
where, for convenience, we have introduced the function
\begin{equation}
	F(A,B) = A q_1(1-z^{q_2})-B q_2 (1-z^{q_1})\,.
\end{equation}
Unfortunately, it is generically not possible to additionally solve $b(q_3)=0$ analytically to obtain a closed expression\footnote{For the interested reader, we record the actual equation that needs to be solved:
\begin{equation*}
b(-q_3)=0\quad\iff\quad	\sum_{\sigma\in S_3} (-1)^{\mathrm{sign}\,\sigma} q_{\sigma(1)}q_{\sigma(2)} z^{q_{\sigma(1)}}\left(1-z^{q_{\sigma(3)}}\right) = 0\,,
\end{equation*}
which is a degree $\mathrm{max}_{i,j=1,2,3}(q_i+q_j)$ polynomial in $z$, which is furthermore fully anti-symmetric in $q_1,q_2,q_3$.} for $z$. Nevertheless, the above solution has a striking feature. Namely, one can show that whenever $|z|=1$ it must again be true that $|\lambda|=1$, as in case (I). 

From the above discussion one concludes that in general
\begin{equation}
	|z|=1\quad\implies\quad |\lambda|=1\,.
\end{equation}
As remarked earlier, this is an interesting observation when compared with the study of a duality between $\lambda$-deformations and $\eta$-deformations \cite{Klimcik_2015}. For completeness, we also remark that for this general set of solutions, the on-shell action again takes the form (see appendix \ref{sec:app_action} for more details)
\begin{equation}
	S_\lambda[g] = i \sum_{q\neq 0}\left[ f(q,\arg z,\arg \lambda) \int_\Sigma d^2t \,\mathrm{Tr}\left(B_{-q}\bar{B}_{+q}\right)\right]\,,
\end{equation}
where $f(q,\arg z,\arg \lambda)$ is a real function of $q$ and the phases of $z$ and $\lambda$ and we have omitted the WZ-term.

\section{Conclusions}

The main purpose of this work has been to uncover a relationship between variations of Hodge structures and solutions to integrable deformed $\sigma$-models. It is based on the identification of the classifying space of Hodge structures with the target space of the $\lambda$-deformed $G/G$ model. One of the motivations to study such a relationship is the importance of the period mapping in string compactifications, for it captures important features of the effective theory with regards to moduli stabilization and the gauge couplings. In contrast, the study of integrable deformed $\sigma$-models is largely motivated by the AdS/CFT-correspondence and general aspects of integrability. Over the years, this has resulted in the development of a vast landscape of integrable field theories. While there still remains much to understand about the precise extent of our proposed relationship, we believe that the combined tools that have been developed to study these two very different fields may yield new insights into the other and we hope to develop this further in future works. 

To elucidate said relationship, we have first described the classical dynamics of the $\lambda$-deformed $G/G$ model, which describes the propagation of a string on a group manifold in the presence of particular background fields. From this point of view, the $\lambda$-deformation is simply a deformation of these target space fields that preserves the integrable nature of the $G/G$ model, whilst breaking its gauge invariance. More concretely, we have formulated the $\lambda$-deformation at the level of the action and simplified the resulting equations of motion. The upshot is that, for a simply-connected string worldsheet and $\lambda\neq \pm 1$, solutions to the model are described in terms of a group-valued field $g$ and an algebra-valued field $U$, where we recall that the gauge field was determined in terms of $U$ via $\mathbf{A}=\star\,\dd U$. The crucial step, then, was the identification of these two fields with the appropriate object that describes a variation of Hodge structures, namely the bulk charge operator $Q$. Indeed, the eigenspace decomposition of $Q$ determines a Hodge structure, which furthermore defines a variation of Hodge structures if the horizontality condition is satisfied.

Concretely, the relationship between variations of Hodge structures and solutions to the $\lambda$-deformed $G/G$ model is given by identifying the field $g$ as the Weil operator $(-1)^Q$ and similarly identifying $U$ as $(i\mp 1)Q$ and additionally putting $\lambda = \pm i$. In other words, the trajectory of a string propagating on the classifying space of Hodge structures in this particular background of the $\lambda$-deformed $G/G$ model precisely describes a variation of Hodge structures. A curious feature of this solution is that $U$ and the deformation parameter $\lambda$ take complex values. However, given that the on-shell action turns out to be purely imaginary and $|\lambda|=1$, this may have an interpretation as a real-valued solution of the $\eta$-deformation, in view of the Poisson-Lie T-duality between $\lambda$-deformations and $\eta$-deformations \cite{Klimcik_2015}. Continuing on, we extended our analysis to a more general ansatz $g=z^Q$, for $z\in\mathbb{C}$, and found that the resulting solutions are still remarkably constrained in the structure of their charge components. Indeed, for these solutions the charge components of $h^{-1}\dd h$ individually satisfy a generalized set of Nahm's equations for each charge. It would be interesting to understand if these generalized solutions play any role in Hodge theory, or an appropriate generalization thereof.  
 
Moving forward, one can interpret our proposed relationship between variations of Hodge structures and $\lambda$-deformed $G/G$ models in two ways, each with their corresponding implications and opportunities for further research. On the one hand, there is an exciting possibility that known techniques to study integrable models can be applied to further understand aspects of Hodge theory and the period mapping. Indeed, in \cite{Appadu:2017fff} (see also \cite{Destri:1988,Destri:1989,Delduc:2012qb,Delduc:2012vq}) it was suggested that the $\lambda$-deformed principal chiral model can be realized as a generalized spin chain model that can be solved via the algebraic Bethe ansatz, which lies at the heart of the quantum inverse scattering method. However, due to issues of non-ultralocality the exact details and possible generalizations to other $\lambda$-deformations remain non-trivial. Nevertheless, in light of these works it is conceivable that one may also describe a variation of Hodge structures in terms of a generalized spin chain, which may offer new methods to solve for period mappings. Another interesting avenue is to consider the recent works on D-branes in $\lambda$-deformations \cite{Driezen:2018glg,Sfetsos:2021pcs}, which may give an alternative interpretation of the intricate boundary conditions that arise in the study of asymptotic Hodge theory. Alternatively, by studying the boundary limit of the $\lambda$-deformed $G/G$ model on an $AdS_2$ worldsheet, along the lines of \cite{Beccaria:2020qtk}, one may shed new light on the proposed holographic structure of string moduli spaces \cite{Grimm:2021ikg}. In an upcoming second part of this work we will concern ourselves precisely with these boundary aspects in order to make further contact with asymptotic Hodge theory \cite{Grimm_WZW_II}.

On the other hand, there is the opportunity to use the techniques of asymptotic Hodge theory to study explicit solutions to the $\lambda$-deformed $G/G$ model. Indeed, as we have briefly mentioned in section \ref{subsec:VHS}, the work of Cattani, Kaplan and Schmid \cite{CKS} provides an algorithmic way to obtain solutions to period mappings in asymptotic regions of the moduli space. Furthermore, this algorithm has recently been applied to find explicit solutions to the period mapping for all one-parameter variations of Hodge structures with $D=3$ and $h^{3,0}=1$, corresponding to those of Calabi-Yau type \cite{Grimm:2021ikg}. It would also be interesting to compare the VHS solution for the case $G=\mathrm{Sp}(2,\mathbb{R})=\mathrm{SL}(2,\mathbb{R})$ to solutions found in \cite{Katsinis:2021nfu}. Of course, an important next step would be to generalize our proposed correspondence beyond the one-modulus case. From the Hodge theoretic point of view it is expected that this is very non-trivial, though perhaps the $\sigma$-model point of view offers a natural generalization in terms of a higher-dimensional $\lambda$-deformed model. We hope to return to these questions in future works. 

\subsubsection*{Acknowledgements}

It is a great pleasure to thank Damian van de Heisteeg, Arno Hoefnagels, Eran Palti, Lorenz Schlechter, Christian Schnell and Dirk Schuricht for very useful discussions and correspondence. This research is partly supported by the Dutch Research Council (NWO) via a Start-Up grant and a VICI grant.

\appendix

\section{Properties of $b(q)$}
\label{sec:app_beta}

In this section we present and derive some algebraic properties of the function
\begin{equation}
	\tilde{b}(q) = z^q - \frac{1-\alpha q\lambda^{-1}}{1-\alpha q}\,.
\end{equation}
Note that this differs from the function $b(q)$ in \eqref{eq:beta} by a factor of $(1-\alpha q)/(\lambda^{-1}-z^q)$. However, since we assume $\alpha q\neq 1$ and $\lambda\neq z^{-q}$, and are only interested in the zeroes of $b(q)$, it suffices to consider $\tilde{b}(q)$. This will turn out to be slightly more convenient. As discussed in the main text, we additionally exclude the following values of the parameters: $z=0,z^q=1$, $\alpha=0$, $\lambda=\pm 1$. 

We are interested in the case where $\tilde{b}(q_1)=\tilde{b}(q_2)=0$ for two distinct integers $q_1,q_2$. One can solve this algebraically to find
\begin{align}\label{eq:sol_alpha_lambda}
	\tilde{b}(q_1)=\tilde{b}(q_2)=0\quad\iff\quad \alpha = -\frac{F(z^{q_1},z^{q_2})}{F(q_2,q_1)}\,,\qquad \lambda  = \frac{F(z^{q_1},z^{q_2})}{F(1,1)}\,,
\end{align}
where, for convenience, we have introduced the function
\begin{equation}
	F(A,B) = A q_1(1-z^{q_2})-B q_2 (1-z^{q_1})\,.
\end{equation}
Using this result, we will now provide a proof of the properties (2b) and (2c), see also \eqref{eq:2b} and \eqref{eq:2c}.

\subsubsection*{Property (2b)}
First, we prove property (2b), which states that
\begin{equation}\label{eq:prop2b}
	1+q_1\alpha\mu = 0\quad\iff\quad b(q_1-q_2)=0\,,
\end{equation}
To this end, one computes both the left-hand side and right-hand side, with $\alpha$ and $\lambda$ given by \eqref{eq:sol_alpha_lambda}. For the right-hand side, this yields
\begin{equation}
	\tilde{b}(q_1-q_2) = z^{-q_2}\frac{\left[z^{q_1}q_1^2(1-z^{q_2})^2-2q_1 q_2(1-z^{q_2})^2+q_2^2(1-z^{q_1})(z^{q_1}-z^{2q_2})\right]}{F(2q_2-q_1, q_2)}
\end{equation}
while the left-hand side becomes
\begin{equation}
	1+q_1\alpha\mu = \frac{\left[z^{q_1}q_1^2(1-z^{q_2})^2-2q_1 q_2(1-z^{q_2})^2+q_2^2(1-z^{q_1})(z^{q_1}-z^{2q_2}) \right]}{q_2(z^{q_1}-z^{q_2})F(1+z^{q_1}, 1+z^{q_2})}\,.
\end{equation}
Indeed, one sees that both the left-hand side and right-hand side are proportional to the same expression (in square brackets), hence we have proven \eqref{eq:prop2b}.

\subsubsection*{Property (2c)}
Finally, we prove property (2c), which states that
\begin{equation}\label{eq:prop2c}
	1+(q_1+q_2)\alpha\mu = 0\quad\iff\quad b(q_1+q_2)=0\,.
\end{equation}
To this end, one first computes the right-hand side, which yields
\begin{equation}
	\tilde{b}(q_1+q_2) = -\frac{q_1^2z^{q_1}\left(1-z^{q_2}\right)^2-q_2^2z^{q_2}\left(1-z^{q_1}\right)^2}{F(q_1,q_2)}\,,
\end{equation}
while the left-hand side is equal to
\begin{equation}
	1+(q_1+q_2)\alpha\mu = \frac{q_1-q_2}{q_1q_2(z^{q_1}-z^{q_2})}\times \frac{q_1^2z^{q_1}\left(1-z^{q_2}\right)^2-q_2^2z^{q_2}\left(1-z^{q_1}\right)^2}{F(1+z^{q_1},1+z^{q_2})}\,.
\end{equation}
Again, one sees that the left-hand side and right-hand side are proportional to the same factor, hence 
\eqref{eq:prop2c} follows. 

\section{Equation of Motion of $g$}\label{sec:app_eomg}

In this section we present some explicit computations regarding the equation of motion of $g$, i.e.
\begin{equation}
	\partial\bar{A}=\mu[A,\bar{A}]\,.
\end{equation}
Using the relations \eqref{eq:g_to_B} and \eqref{eq:A_to_B} one readily computes
\begin{align}
	\partial\bar{A} &= \alpha \sum_{q\neq 0} q\, h\left( \partial\bar{B}_q + [h^{-1}\partial h, \bar{B}_q]\right)h^{-1} = \alpha \sum_{q\neq 0} q\,h \left( \partial\bar{B}_q + \sum_{q'} [B_{q'},\bar{B}_q]\right) h^{-1}\,.
\end{align}
It follows that
\begin{align}
	\partial\bar{A} = \mu[A,\bar{A}]\quad &\iff \quad \alpha \sum_{q\neq 0} q\, \left( \partial\bar{B}_q + \sum_{q'} [B_{q'},\bar{B}_q]\right) = \alpha^2 \mu \sum_{q\neq 0} \sum_{q'\neq 0} q q' [B_{q'},\bar{B}_q]\,,\\
	&\iff\quad  0= \sum_{q\neq 0}q\left[ \partial\bar{B}_q+ \sum_{q'}(1-q'\alpha\mu)[B_{q'},\bar{B}_q]\right]\,.
\end{align}
Projecting the final equation onto $q^*$ modes and relabeling $q'\rightarrow q$ gives
\begin{align}\label{eq:Bbar}
	 0=q^*\left(\partial\bar{B}_{q^*}+[B_0,\bar{B}_{q^*}]\right) + \sum_{q\neq 0,q^*}q (1+(q-q^*)\alpha\mu)[B_{q^*-q}, \bar{B}_{q}]\,.
\end{align}
Using the Bianchi identity
\begin{align}
	0 &= (\partial\bar{B}-\bar{\partial}B+[B,\bar{B}])_{q^*}\\
	&=\left(\partial\bar{B}_{q^*}+[B_0,\bar{B}_{q^*}]\right)-\left(\bar{\partial}B_{q^*}+[\bar{B}_0,B_{q^*}]\right)+\sum_{q\neq 0,q^*} [B_{q^*-q},\bar{B}_q]\,,
\end{align}
and subtracting it from \eqref{eq:Bbar} yields the similar equation
\begin{equation}\label{eq:B}
	0=q^*\left(\bar{\partial}B_{q^*}+[\bar{B}_0,B_{q^*}]\right)+\sum_{q\neq 0,q^*}(q-q^*)(1+q\alpha\mu)[B_{q^*-q}, \bar{B}_{q}]\,.
\end{equation}
In the main text, equations \eqref{eq:Bbar} and \eqref{eq:B} are studied further.

\section{Evaluating the On-shell Action}\label{sec:app_action}

By evaluating the action on the constraints \eqref{eq:Abar_ginvdbarg} and \eqref{eq:A_ginvdg} one obtains the on-shell action
\begin{equation}
	S_\lambda[g] =\frac{k}{\pi}\int_\Sigma d^2t\,\mathrm{Tr}\left(g^{-1}\partial g\left[\frac{1}{2}+\frac{\lambda\,\mathrm{Ad}_g}{1-\lambda \,\mathrm{Ad}_g} \right]g^{-1}\bar{\partial}g \right)+ S_{\mathrm{WZ}}[g]\,.
\end{equation}
By recalling that
\begin{equation}
	g^{-1}\bar{\partial}g = -\sum_{q\neq 0} (1-z^{-q})h \bar{B}_q h^{-1}\,,
\end{equation}
one readily sees that
\begin{equation}
	\frac{\lambda\,\mathrm{Ad}_g}{1-\lambda\, \mathrm{Ad}_g} g^{-1}\bar{\partial}g = -\sum_{q\neq 0}(1-z^{-q}) \frac{\lambda z^{q}}{\lambda-z^{q}}h \bar{B}_q h^{-1}\,.
\end{equation}
Furthermore, since for non-zero $q,q'$
\begin{equation}
	\mathrm{Tr}\left(\mathcal{O}_q\cdot\mathcal{O}_{q'}\right)=\frac{1}{q'}\mathrm{Tr}\left(\mathcal{O}_q\cdot[Q_{\mathrm{ref}},\mathcal{O}_{q'}]\right) = -\frac{1}{q'}\mathrm{Tr}\left([Q_{\mathrm{ref}},\mathcal{O}_q]\cdot\mathcal{O}_{q'}\right)=-\frac{q}{q'}\mathrm{Tr}\left(\mathcal{O}_q\cdot\mathcal{O}_{q'}\right)\,,
\end{equation}
it follows that $\mathrm{Tr}\left(\mathcal{O}_q\cdot\mathcal{O}_{q'}\right)$ is only non-zero for $q'=-q$. Therefore the on-shell action simplifies to
\begin{equation}
	S_\lambda[g] =\frac{k}{\pi}\sum_{q\neq 0}\left[(1-z^{q})(1-z^{-q})\left(\frac{1}{2}-\frac{1}{1-z^{-q}/\lambda}\right)\int_\Sigma d^2t\,\mathrm{Tr}\left(B_{-q}\bar{B}_{+q} \right)\right] +S_{\mathrm{WZ}}[g]
\end{equation}
The term inside the integral is positive definite, while the coefficients satisfy
\begin{equation}
	\mathrm{Re}\left[\frac{1}{2}-\frac{1}{1-z^{-q}/\lambda}\right] = 0 \quad \iff \quad |\lambda| = |z|^{-q}. 
\end{equation}
In particular, when $|z|=|\lambda|=1$ this is satisfied for all $q$, hence the action is purely imaginary-valued. Concretely, introducing the angles
\begin{equation}
	z = e^{2i\theta}\,,\qquad \lambda = e^{2i\phi}\,,
\end{equation} 
one finds
\begin{equation}
	S_\lambda[g] =\frac{2i k}{\pi}\sum_{q\neq 0}\left[\sin^2(q\theta)\cot(q\theta+\phi)\int_\Sigma d^2t\,\mathrm{Tr}\left(B_{-q}\bar{B}_{+q} \right)\right] +S_{\mathrm{WZ}}[g]
\end{equation}
Moreover, for the VHS solution one has $\theta=\pi/2$ and $\phi = \pm \pi/4$ and the summand only runs over $q=1$. In that case, the on-shell action reduces to
\begin{equation}
	S_\lambda[g] = \mp\frac{2i k}{\pi}\int_\Sigma d^2t\,\mathrm{Tr}\left(B_{-1}\bar{B}_{+1} \right) +S_{\mathrm{WZ}}[g]\,.
\end{equation}

\bibliographystyle{jhep}
\bibliography{references}

\end{document}